%% file: cna-lock.tex
\pdfoutput=1

\documentclass[sigplan,10pt,nonacm=true]{acmart}

\usepackage{microtype}

\usepackage{listings}

\usepackage{graphicx}
\usepackage{makecell}
\usepackage{multirow}
\usepackage{caption}
\usepackage{subfig}
\usepackage[export]{adjustbox}
\usepackage{url}
\usepackage[htt]{hyphenat} 

\usepackage[iso,american,cleanlook]{isodate}
\hypersetup{hidelinks=false,colorlinks=true,linkcolor=blue,urlcolor=blue,citecolor=blue,anchorcolor=blue}
\include{defs}

\copyrightyear{2019} 
\acmYear{2019} 
\setcopyright{acmcopyright}
\acmConference[EuroSys '19]{Fourteenth EuroSys Conference 2019}{March 25--28, 2019}{Dresden, Germany}
\acmBooktitle{Fourteenth EuroSys Conference 2019 (EuroSys '19), March 25--28, 2019, Dresden, Germany}
\acmPrice{15.00}
\acmDOI{10.1145/3302424.3303984}
\acmISBN{978-1-4503-6281-8/19/03}

\def\ContinueLineNumber{\lstset{firstnumber=last}}

\makeatletter
\lst@Key{countblanklines}{true}[t]%
    {\lstKV@SetIf{#1}\lst@ifcountblanklines}

\lst@AddToHook{OnEmptyLine}{%
    \lst@ifnumberblanklines\else%
       \lst@ifcountblanklines\else%
         \advance\c@lstnumber-\@ne\relax%
       \fi%
    \fi}
\makeatother

\lstdefinestyle{numbers}
{numbers=left, stepnumber=1, numberstyle=\tiny, numbersep=10pt}
\lstdefinestyle{nonumbers}
{numbers=none}

\title{Compact NUMA-aware Locks}

\author{Dave Dice}
\affiliation{Oracle Labs}
\email{first.last@oracle.com}
\author{Alex Kogan}
\affiliation{Oracle Labs}
\email{first.last@oracle.com}

\begin{document}

\begin{abstract}
Modern multi-socket architectures exhibit non-uniform memory access (NUMA) behavior, where access by a core to data cached locally
on a socket is much faster than access to data cached on a remote socket.
Prior work offers several efficient NUMA-aware locks that exploit this behavior by keeping the lock ownership on the same socket, thus reducing
remote cache misses and inter-socket communication.
Virtually all those locks, however, are hierarchical in their nature, thus requiring space proportional to the number of sockets.
The increased memory cost renders NUMA-aware locks unsuitable for systems that are conscious to space requirements of their synchronization constructs,
with the Linux kernel being the chief example.

In this work, we present a compact NUMA-aware lock that requires only one word of memory, regardless of the number of sockets in the underlying machine.
The new lock is a variant of an efficient (NUMA-oblivious) MCS lock, and inherits its performant features, such as local spinning and a single atomic instruction in
the acquisition path.
Unlike MCS, the new lock organizes waiting threads in two queues, one composed of threads running on the same socket 
as the current lock holder, and another composed of threads running on a different socket(s).

We implemented the new lock in user-space as well as integrated it in the Linux kernel's \emph{qspinlock}, one of the major synchronization constructs in the kernel.
Our evaluation using both user-space and kernel benchmarks shows that the new lock 
has a single-thread performance of MCS, but significantly outperforms the latter under contention,
achieving a similar level of performance when compared to other, state-of-the-art NUMA-aware locks that require substantially more space. 

 \end{abstract}


%

\maketitle

\thispagestyle{fancy}

\section{Introduction}

\remove{
Outline:
1. Locks remains perhaps the most popular synchronization technique.
2. Modern architectures feature multiple sockets. To be efficient, synchronization algorithms have to be NUMA-aware.
3. Many NUMA-aware locks exist. All of them require space proportional to the number of sockets.
4. This work proposes a compact NUMA-aware lock that requires one word, essentially the same amount of memory as non-NUMA-aware lock.
The lock is an adaptation of the popular MCS lock.
5. Evaluation of multiple benchmarks shows the benefit of the new lock.
}

Locks are used by concurrently running processes (or threads) to acquire exclusive access to shared data.
Since their invention in the mid sixties~\cite{Dij65}, 
locks remain the topic of extensive research 
and the most popular synchronization technique in parallel software.
Prior research has shown that the performance of such software often depends directly on the 
efficiency of the locks it employs~\cite{EE10, Dice17, GLQ16}
\footnote{A version of this paper appears in EuroSys 2019 : \url{https://doi.org/10.1145/3302424.3303984}}. 

The evolution of locks is tightly coupled with the evolution of computing architectures.
Modern architectures feature an increasing number of nodes (or sockets), each comprising of a locally attached memory, a fast local cache
and multiple processing units (or cores).
Accesses by a core to a local memory or local cache are significantly faster than accesses to a remote memory or cache lines residing on another node~\cite{RH03}, 
characteristic known as NUMA (Non-Uniform Memory Access).
As a result, researchers have proposed multiple designs for NUMA-aware locks, 
which try to keep the lock ownership within the same socket~\cite{CMC16, RH03, DMS15, CFM15, LNS06, DMS11, KMK17}.
This approach decreases remote cache misses and the associated inter-socket communication, as
it increases the chance that the lock data, as well as the shared data subsequently accessed in a critical section, will be cached locally to the socket
on which a lock holder is running.

While performance evaluations show that NUMA-aware locks perform substantially better than their NUMA-oblivious counter-parts,
one particular issue hampers the adoption of the former in practice.
Specifically, while NUMA-oblivious locks can be implemented using a single memory word (or even a bit), 
virtually all NUMA-aware locks are hierarchical in their nature, 
built of a set of \emph{local} (typically, per-socket) locks each mediating 
threads running on the same socket and a \emph{global} lock synchronizing 
threads holding a local lock~\cite{CMC16, DMS11, DMS15, CFM15, LNS06, KMK17}.

The hierarchical structure of NUMA-aware locks is problematic for three reasons.
First, even when the lock is uncontended, a thread still has to perform multiple atomic operations to acquire multiple 
low-level locks of the hierarchy before it can enter a critical section.
This often results in suboptimal single-thread performance when compared to efficient NUMA-oblivious locks.
Second, to ensure portability, NUMA-aware locks have to be initialized dynamically as the number of sockets 
of the underlying system is unknown until the run time.
Beyond the inability to allocate lock instances statically, such initialization requires querying the topology of the underlying system,
which in fact hinders the portability as no standard APIs for those queries exist.
Third, and perhaps most importantly in the context of this work, hierarchical NUMA-aware locks 
require space proportional to the number of sockets.
Making the matter even worse, each low-level lock of the hierarchy has to be placed on a separate cache line 
in order to achieve scalability and avoid false sharing.
In certain environments, such an increase in the space requirement is prohibitively expensive.
One example is systems that feature numerous (e.g., millions) of locks, such as database systems or an operating system kernel.
The Linux kernel, for instance, strictly limits the size of its spin lock to $4$ bytes.
Among many use cases, this lock is embedded in the inode (index node) and page structures, 
which represent, respectively, information about each file and each physical page frame on a system~\cite{Bue14}.
As a result, any increase to the size of the lock would be unacceptable~\cite{linux-locks, linux-page-struct}.
As Bueso notes~\cite{Bue14}, ``the more files or memory present, the more instances of these structures are handled by the kernel. 
It is not uncommon to see machines with tens of millions of cached inodes'', so
even a minor increase in the size of the inode structure (e.g., due to the increase in the size of the lock)
would be ``enough to go from having a well-balanced workload to not being able to fit the working set of inodes in memory''.
He also urges implementers to ``always keep in mind the size of the locking primitives''~\cite{Bue14}.
Another example where the size of the lock is important is in concurrent data structures, such as linked lists or binary search trees,
that use a lock per node or entry~\cite{BCO10, CGR13, HHL05}.
Contention on such locks may arise when the workload is skewed, and a small set of nodes becomes heavily accessed.
As Bronson at el.\ note, when a scalable lock is striped across multiple cache lines to avoid contention in the coherence fabric,
it is ``prohibitively expensive to store a separate lock per node''~\cite{BCO10}.

This work proposes a compact NUMA-aware lock, called CNA, that requires one word only, regardless of the number of sockets of the underlying machine.
Moreover, CNA requires only one atomic instruction per lock acquisition.
The CNA lock is a variant of the popular and highly efficient MCS lock~\cite{MCS91}, in which threads waiting for the lock form a queue and spin on a local cache 
line until the lock becomes available to them.
CNA attempts to pass the lock to a successor running on the same socket, rather than to a thread that happens to be next in the queue.
While looking for the successor, the lock holder moves waiting threads running on a different socket(s) to a separate, secondary queue, so they do not interfere
in subsequent lock handovers.
Threads in the secondary queue are moved back to the primary queue under one of the two conditions: 
(a) when the main queue does not have any waiting threads running on the same socket as the current lock holder, or
(b) after a certain number of local handovers.
The latter condition is to ensure long-term fairness and avoid starvation (of threads in the secondary queue).

\remove{
To avoid deadlock, a pointer to the secondary queue has to be known to each lock holder 
(so it can pass the lock ownership to the thread at the head of that queue if the ``main'' queue becomes empty).
A natural place to store this pointer would be in the lock structure, but that would increase the space of the latter (which stores  only
a pointer to the tail of the ``main'' queue).
We solve this issue by effectively reusing the field through which the lock ownership is passed (same field on which threads waiting for the lock spin).
}

We implemented the CNA lock as a stand-alone dynamically linked library conforming to the POSIX pthread API.
We also modified the Linux kernel spin lock implementation (\emph{qspinlock}) to use CNA.\footnote{The patch is available at \url{https://lwn.net/Articles/778235}.}
We evaluated the CNA lock using a user-space microbenchmark, multiple real applications and several kernel
microbenchmarks.
The results show that, unlike many NUMA-aware locks, CNA does not introduce any overhead in 
single-thread runs over the MCS lock.
At the same time, it significantly outperforms MCS under contention (typically, by about $40\%$ or more on a two-socket 
system and by about $100\%$ or more on a four-socket system), achieving a similar
level of performance compared to other, state-of-the-art NUMA-aware locks that require substantially more space.

The rest of the paper is organized as following.
We survey the related work in \secref{sec:related}.
\secref{sec:qspinlock} provides the details of the current Linux kernel spin lock implementation.
We present the CNA design overview in \secref{sec:design} followed by implementation details in \secref{sec:implementation}
and possible optimizations in \secref{sec:opts}.
The results of an extensive evaluation are given in \secref{sec:evaluation}, and we conclude in \secref{sec:conclusion}.

\section{Related Work}
\seclabel{sec:related}

A test-and-set lock~\cite{And90} is one of the simplest spin locks. 
To acquire this lock, a thread repeatedly executes an atomic test-and-set instruction
until it returns a value indicating that the state of the lock has been changed from
unlocked to locked.
While this lock can be implemented with just one memory word (or even bit), it employs
\emph{global spinning}, that is all threads trying to acquire the lock spin on the same memory location.
This leads to excessive coherence traffic during lock handovers.
Furthermore, this lock does not provide any fairness guarantees, as a thread that just released the lock 
can return and bypass threads that have been waiting for the lock for a long time.

Queue locks address those challenges by organizing threads waiting for the lock in a FIFO queue.
In MCS~\cite{MCS91}, one of the most popular queue locks, the shared state of the lock consists of a pointer to a tail of the queue.
Each thread has a record (queue node) that it inserts into the queue (by atomically swapping the tail), 
and then spins locally on a flag inside its record, which will be set by its predecessor when the latter unlocks the lock.
In general, queue spin locks provide faster lock handover under contention compared to locks with global spinning~\cite{And90}.
Those locks, however, are not friendly to NUMA systems, since they cause increased coherence 
traffic as the lock data (and the data accessed in the critical section) can migrate from one socket to another with every lock
handover.

In order to keep the lock on the same socket, 
Radovic and Hagersten propose a hierarchical backoff lock (HBO)~\cite{RH03}, 
which requires only one word of memory.
The idea is to use that word to store the socket number of the lock holder.
When a thread finds the lock unavailable, it sets its back-off to a small value if it runs on the same node as the lock holder and to a larger value if 
it runs on a different socket.
This approach, however, poses the same challenges as spin locks with global spinning~\cite{LNS06}.
In particular, threads spinning on other sockets may be starved, and they create contention on the lock state even if they check the state less frequently.
(To mitigate those issues, the authors considered several extensions of the basic HBO algorithm (using some extra words of memory), 
yet they do not eliminate them completely~\cite{RH03}).
Moreover, backoff timeouts require tuning for optimal performance, and this tuning is known to be challenging~\cite{JAS09, LA93}.

Subsequent designs of NUMA-aware locks use a hierarchy of synchronization constructs (e.g., queue locks), with per-socket constructs
serving for intra-socket synchronization and a construct at the top of the hierarchy synchronizing between threads on different sockets.
This idea was realized by Luchangco et al.~\cite{LNS06} and later by Dice et al.~\cite{DMS11}.
Dice et al.~\cite{DMS15} generalize this approach in the \emph{Lock Cohorting} technique that constructs a NUMA-aware lock out of
any two spin locks L and G that have certain properties.
Chabbi et al.~\cite{CFM15} generalize the Lock Cohorting technique further to multiple levels of hierarchy, addressing architectures with a deeper NUMA hierarchy.
They present HMCS, a hierarchical MCS lock, in which each level of the hierarchy is protected by an MCS lock.

Hierarchical NUMA-aware locks have a memory footprint with the size 
proportional to the number of sockets (or more precisely, 
to the number of nodes in the hierarchy tree).
Note that in order to achieve scalability and avoid false sharing,  
per-socket constructs have to placed on different cache lines, inflating the 
size of the lock further.
Besides, hierarchical locks tend to perform poorly under no or light contention, 
since a lock operation involves multiple atomic instruction (to acquire multiple locks).
To address some of those challenges, Kashyap et al.~\cite{KMK17} present a hierarchical NUMA-aware CST lock, which defers 
the allocation of per-socket locks until the moment each of those locks is accessed for the first time.
This is useful in environments where threads are restricted to run on a subset of sockets, yet the memory footprint of the CST lock grows linearly with the 
number of sockets in the general case.
A different angle is taken by Chabbi and Mellor-Crummey~\cite{CMC16}, where 
the authors augment the HMCS lock~\cite{CFM15} with a fast path.
This path enables threads to bypass 
multiple levels of the hierarchy in HMCS when there is no contention and compete directly for the lock at the top of the hierarchy.
This results in a contention-conscious hierarchical lock that performs similarly to MCS under no contention, 
but increases the memory cost of the HMCS lock even further, as it has to maintain metadata that 
helps threads to estimate the current level of contention.

In a different, but highly related area, Dice~\cite{Dice17} explored the scalability collapse phenomenon, 
in which the throughput of a system drops abruptly due to lock contention.
He suggests to modify the admission policy of a lock, limiting the number of distinct threads circulating though the lock.
In the case of MCS, excessive threads are removed from the MCS lock queue and placed into a separate list.
While the resulting lock (called MCSCR~\cite{Dice17}) is shown to perform well under contention, and in particular on over-subscribed systems,
it is NUMA-oblivious and uses multiple words of memory (to keep track of the multiple queues/lists).
As a future direction, Dice mentions a possibility for constructing MCSCRN, a NUMA-aware version of MCSCR.
In addition to the state of MCSCR, MCSCRN is conceived to also include two extra fields: the identity of the current preferred socket and a pointer to the list 
of ``remote'' threads running on a different socket(s)~\cite{Dice17}.

\section{Background (Linux Kernel Spin Lock)}
\seclabel{sec:qspinlock}
Being one of the major synchronization constructs in the Linux kernel and having 
``a great deal of influence over the safety and performance of the kernel''~\cite{Cor08},
the spin lock implementation has evolved over the course of the years.
\remove{
Originally, it was implemented as a variant of a test-and-set lock, in which 
a thread atomically decrements the value stored in the lock and checks to see whether the result is zero.
If so, it has a lock; otherwise, a negative number indicated the number of threads waiting for the lock.
Those threads have to spin in a tight loop waiting for the lock value to become positive, and retry the acquisition.
The lock owner release the lock by setting it atomically to $1$.

As hardware systems have evolved to include multiple sockets, kernel maintainers realized that such an implementation
can be grossly unfair and cause starvation.
In response to that, around 2008 a new implementation has been put into place, which is based on ticket locks~\cite{Cor08}.
Ticket locks are split into two fields, called \texttt{next} and \texttt{owner}, packed into one word of memory and initially set to $0$.
A thread acquiring the lock atomically reads the value of both fields and increments the value of \texttt{next}.
If \texttt{next} and \texttt{owner} are equal before the increment, the thread has the lock.
Otherwise, it spins and waits until the value of \texttt{owner} is incremented to match the value it read from \texttt{next}.
When a thread exist a critical section, it simply increments \texttt{owner} (albeit it requires an atomic operation).
}
The current implementation of spin locks uses a multi-path approach, with a fast path
implemented as a test-and-set lock and a slow path implemented as an MCS lock~\cite{Long13}.
More specifically, a four-byte lock word is divided into three parts: the lock value, the pending bit and the queue tail.
A thread acquiring a spin lock tries first to flip atomically the value of the lock word from $0$ to $1$, and if successful, it has the lock.
Otherwise, it checks whether there is any contention on the lock.
The contention is indicated by any other bit in the lock word being set (the pending bit or the bits of the queue tail).
If there is no contention, the thread attempts to set atomically the pending bit, and if successful, spins and waits until the 
current lock holder releases the lock. 
If contention is detected, the thread switches to a slow path, in which it enters an MCS queue.

Once a thread $t$ enters the MCS queue (by atomically swapping the queue tail in the lock word with an encoded pointer to its queue node), 
it waits until it reaches the head of the queue.
It happens if $t$ enters an empty queue, or when $t$'s predecessor in the queue writes into a flag in $t$'s queue node on which $t$ spins.
At that point, $t$ waits for the lock holder and a thread spinning on the pending bit (if such a thread exists) to go away.
This in turn happens when $t$ finds both the lock value and the pending bit being clear.
At that time, $t$ claims the lock (by setting non-atomically the lock value to $1$), and writes into the flag of its
successor $s$ in the MCS queue (if such successor exists), notifying the latter that now $s$ became to be the thread at the top of the MCS queue.

A thread releasing a spin lock simply sets the lock value to $0$.
It is interesting to note that unlike the original MCS lock~\cite{MCS91}, this design avoids the 
need to carry a queue node from lock to unlock, since the release of the spin lock does not involve queue nodes.
Furthermore, the Linux kernel limits the number of contexts that can nest and in which a spin lock can be acquired (the limit is four).
This allows all queue nodes to be statically preallocated and at the same time, it enables space and time-efficient 
encoding of the tail pointer to make the entire lock word fit into four bytes~\cite{Long13}.

The CNA lock is used to replace the slow path of the kernel spin lock, leaving the
fast path as well as the unlock procedure intact.
Thus, the change to the kernel is minimal and localized to just a few files.
The design of the CNA lock is detailed next.

\section{Design Overview}
\seclabel{sec:design}
The CNA lock can be seen as a variant of the MCS lock with a few notable differences.
MCS organizes threads waiting to acquire the lock in one queue.
CNA organizes threads into two queues, the ``main'' queue composed of threads running on the same 
socket as the lock holder, and a ``secondary'' queue composed of threads
running on a different socket (or sockets).
When a thread attempts to acquire the CNA lock, it always joins the main queue first;
it might be moved to the secondary queue by a lock holder running on a different socket as detailed below 
and exemplified in \figref{fig:scheme}.

\begin{figure*}
\centering
\subfloat[][The main queue consists of six threads, with $t1$, $t4$ and $t5$ running on socket $0$ and the rest running on socket $1$.
Thread $t1$ has the lock. The secondary queue is empty. ]{\includegraphics[width=1\linewidth, Clip=0 5.5cm 5.5cm 0.5cm]{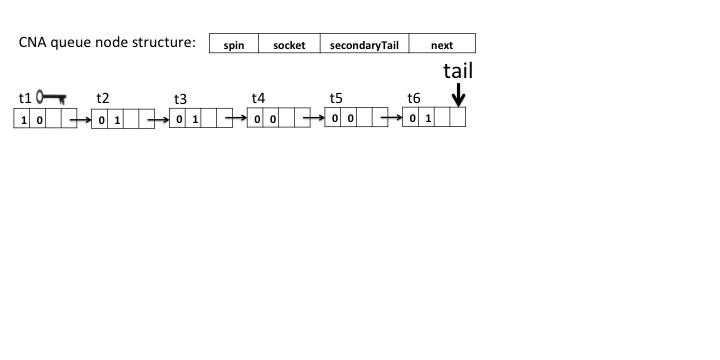}}\\
\subfloat[][$t1$ exits its critical section, traverses the main queue and finds that $t4$ runs on the same socket. 
Therefore, $t1$ moves $t2$ and $t3$ into the secondary queue, setting the \code{secondaryTail} field in $t2$'s node to $t3$.
Then, $t1$ passes the lock to $t4$ by writing the pointer to the head of the secondary queue into $t4$'s \code{spin} field.]{\includegraphics[width=1\linewidth, Clip=0 6.5cm 5.5cm 0.75cm]{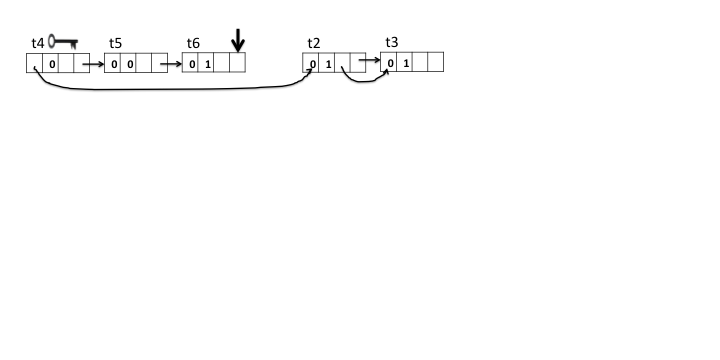}}\\
\subfloat[][$t1$ returns and enters the main queue.]{\includegraphics[width=1\linewidth, Clip=0 6.5cm 5.5cm 0.75cm]{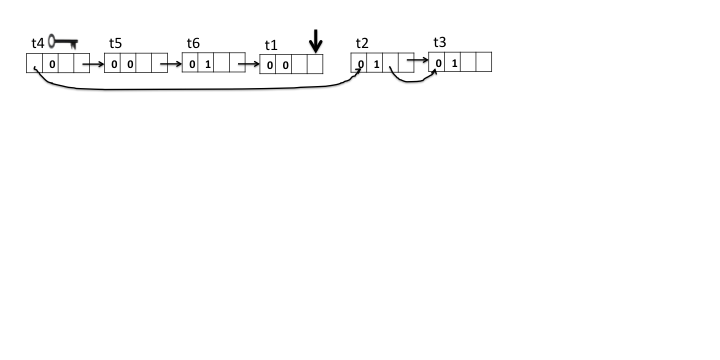}}\\
\subfloat[][When $t4$ exits its critical section, it discovers that the next thread in the main queue ($t5$) runs on the same socket. $t4$ passes the lock to $t5$ by simply copying the value in $t4$'s \code{spin} field into $t5$'s \code{spin} field.]{\includegraphics[width=1\linewidth, Clip=0 6.5cm 5.5cm 0.5cm]{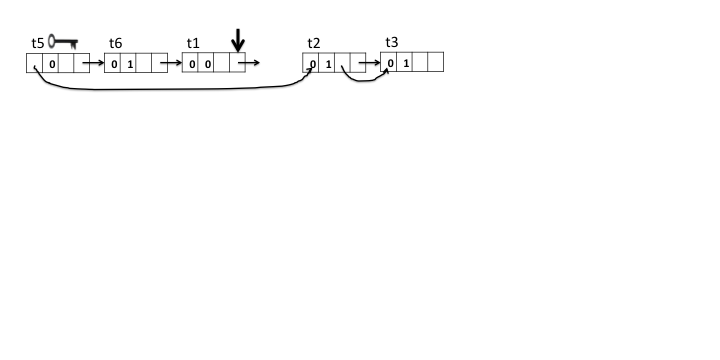}}\\
\subfloat[][$t7$ running on socket $1$ arrives and enters the main queue.]{\includegraphics[width=1\linewidth, Clip=0 6.5cm 5.5cm 0.75cm]{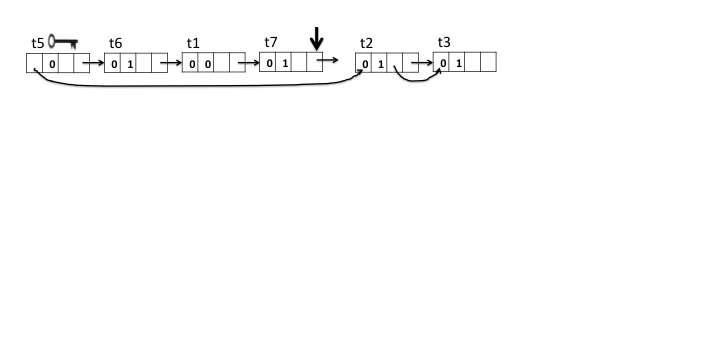}}\\
\subfloat[][$t5$ exits its critical section, moves $t6$ into the end of the secondary queue (and updates the \code{secondaryTail} field in $t2$'s node), and passes the lock to $t1$.]{\includegraphics[width=1\linewidth, Clip=0 6.5cm 5.5cm 0.75cm]{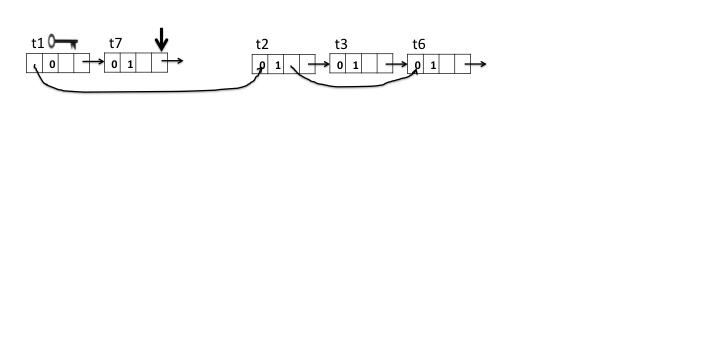}}\\
\subfloat[][$t1$ exits its critical section and finds no threads on socket $0$ in the main queue. Thus, $t1$ moves nodes from the secondary queue back to the main one, putting them before its successor $t7$, and passes the lock to $t2$.]{\includegraphics[width=1\linewidth, Clip=0 6.5cm 5.5cm 0.75cm]{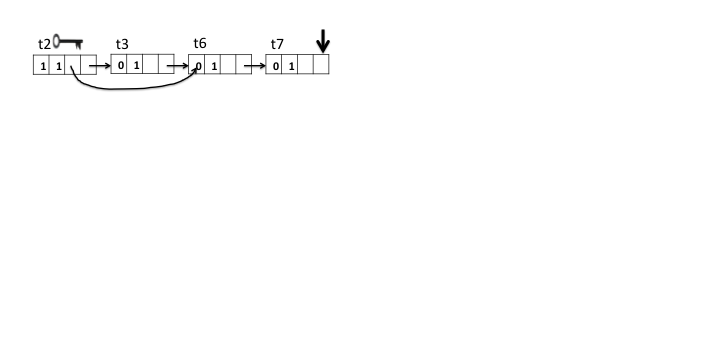}}
\caption{A running example for CNA lock handovers on a $2$-socket machine. Empty cells represent \code{NULL}  pointers.}
\figlabel{fig:scheme}
\end{figure*}

Despite managing two queues of waiting threads, 
the shared state of the CNA lock is comprised of one single word, \code{tail}, which is a pointer to the tail of the main queue.
Like the MCS lock, CNA acquisition requires an auxiliary node, which a thread $t$ inserts into the main queue by atomically swapping 
the \code{tail} pointer with a pointer to its node.
(This is the only atomic instruction performed in the lock acquisition procedure.)
The auxiliary node contains the \code{next} field pointing to the next node in the (main or secondary) queue, and the \code{spin} field on 
which $t$ would spin waiting for it to change (e.g., from $0$ to $1$).
This change will happen when $t$'s predecessor in the queue exits its critical section and 
executes the unlock function, passing the ownership of the lock to $t$.

While the lock function in CNA is almost identical to that in MCS, 
the unlock function is where the CNA lock fundamentally differs.
Instead of simply passing the lock to its successor, the current lock holder $h$ would traverse the main queue and look for a thread
running on the same socket (the socket of each thread is recorded in its node).
If it finds one, say a thread $h^\prime$, it moves all threads (or more precisely, all nodes) between $h$ and $h^\prime$ 
from the main queue to the secondary queue, and passes the ownership to $h^\prime$ (by changing the \code{spin} field in the node belonging to $h^\prime$).
See~\figref{fig:scheme}~(b) for an example.

Note that $h$ needs to pass the pointer to the head of the secondary queue to $h^\prime$ (which the latter will have to pass to its successor, and so on).
This is needed so that $h^\prime$ (or one of its successors) would be able to pass the lock to a thread in the secondary queue, e.g., if the main queue becomes empty.
One natural place to record this pointer is in the lock structure, but that would increase the lock structure space.
Another alternative is to record the pointer in the $h^\prime$'s queue node in a separate field, and then have $h^\prime$ copy the pointer to its successor, and so on.
That would require an extra store instruction (and, potentially, an extra cache miss) during lock handover.
Our approach is to reuse the \code{spin} field for that purpose, that is, instead of handing the lock to $h^\prime$ by writing $1$, 
$h$ writes the pointer to the head of the secondary queue into $h^\prime$'s \code{spin}.
We assume here that a valid pointer cannot have the value $1$, which is true for most systems.

We are left to discuss when threads in the secondary queue would be able to return to the main queue and acquire the lock.
That would happen under one of the two conditions.
First, if the current lock holder $h$ cannot find a thread in the main queue running on the same socket, 
it would place the last node in the secondary queue (i.e., the secondary tail) before $h$'s immediate successor in the main queue and pass the lock 
to the first node in the secondary queue (cf.~\figref{fig:scheme}~(g)).
This would effectively empty the secondary queue.
Note that in order to find the secondary tail, one could scan the secondary queue starting from its head, but that would be inefficient if the queue gets long.
As an optimization, we store the pointer to the tail of the secondary queue in the node belonging to the head of that queue.
We also use (and update) that pointer when moving nodes from the main queue (in the unlock function) into the (non-empty) secondary queue.

The second condition for moving nodes from the secondary to the main queue deals with the long-term fairness guarantees.
As described so far, if threads running on the same socket repeatedly try to acquire the lock (and enter the main queue), 
CNA might starve all other threads running on a different socket (or sockets).
To avoid this case, CNA periodically moves threads from the secondary queue to the main one, effectively passing the lock ownership to
a thread on a different socket.
To this end, we employ a lightweight pseudo-random number generator, and empty the secondary queue with a low (but non-zero) probability.

\section{Implementation Details}
\seclabel{sec:implementation}

\begin{figure}
\begin{lstlisting}[style=nonumbers]
typedef struct cna_node {
  uintptr_t spin;
  int socket;
  struct cna_node *secTail; 
  struct cna_node *next; 
} cna_node_t;

typedef struct {
  cna_node_t *tail;
} cna_lock_t ;

\end{lstlisting}
\caption{Lock and node structures.}
\label{fig:structures}
\end{figure}

\begin{figure}
\begin{lstlisting}[style=numbers]

int cna_lock(cna_lock_t *lock, cna_node_t *me) {
    me->next = 0;
    me->socket = -1;										@\linelabel{lock:3}@
    me->spin = 0;

    /* Add myself to the main queue */
    cna_node_t *tail =  SWAP(&lock->tail, me);					@\linelabel{lock:6}@

    /* No one there? */
    if (!tail) { me->spin = 1; return 0; }							@\linelabel{lock:8}@

    /* Someone there, need to link in */
    me->socket = current_numa_node();						@\linelabel{lock:10}@
    tail->next = me;										@\linelabel{lock:11}@

    /* Wait for the lock to become available */
    while (!me->spin) { CPU_PAUSE(); }						@\linelabel{lock:13}@

    return 0;
}
\end{lstlisting}
\caption{Lock procedure. SWAP stands for an atomic exchange instruction, while CPU\_PAUSE is a no-op used for polite busy waiting.}
\label{fig:lock}
\end{figure}

\ContinueLineNumber
\begin{figure}[t]
\begin{lstlisting}[style=numbers]
void cna_unlock(cna_lock_t *lock, cna_node_t *me) {
  /* Is there a successor in the main queue? */ 
  if (!me->next) {										@\linelabel{unlock:3}@
    /* Is there a node in the secondary queue? */
    if (me->spin == 1) {									@\linelabel{unlock:5}@
      /* If not, try to set tail to NULL, indicating that 
         both main and secondary queues are empty */
      if (CAS(&lock->tail, me, NULL) == me) return;				@\linelabel{unlock:8}@
    } else {												@\linelabel{unlock:9}@
      /* Otherwise, try to set tail to the last node in 
         the secondary queue */
      cna_node_t *secHead = (cna_node_t *)me->spin;			@\linelabel{unlock:11}@
      if (CAS(&lock->tail, me, secHead->secTail) == me) {			@\linelabel{unlock:12}@
        /* If successful, pass the lock to the head of 
           the secondary queue */
        secHead->spin = 1;									@\linelabel{unlock:14}@
        return;											@\linelabel{unlock:15}@
      }
    }

    /* Wait for successor to appear */
    while (me->next == NULL) { CPU_PAUSE(); }				@\linelabel{unlock:19}@
  }													@\linelabel{unlock:20}@

  /* Determine the next lock holder and pass the lock by 			@\linelabel{unlock:21}@
     setting its spin field */
  cna_node_t *succ = NULL;
  if (keep_lock_local() && (succ = find_successor(me))) {
     succ->spin = me->spin;								@\linelabel{unlock:23}@
  } else if (me->spin > 1) {									@\linelabel{unlock:24}@
    succ = (cna_node_t *)me->spin;							@\linelabel{unlock:25}@
    succ->secTail->next = me->next;							@\linelabel{unlock:26}@
    succ->spin = 1;										@\linelabel{unlock:27}@
  } else {												@\linelabel{unlock:28}@
    me->next->spin = 1;									@\linelabel{unlock:29}@
  }
}
\end{lstlisting}
\caption{Unlock procedure. CAS stands for an atomic compare-and-swap instruction.}
\label{fig:unlock}
\end{figure}

\ContinueLineNumber
\begin{figure}[t]
\begin{lstlisting}[style=numbers]
cna_node_t *find_successor(cna_node_t *me) {				@\linelabel{find-succ:1}@
  cna_node_t *next = me->next;								@\linelabel{find-succ:2}@
  int mySocket = me->socket;								@\linelabel{find-succ:3}@
  if (mySocket == -1) mySocket = current_numa_node();			@\linelabel{find-succ:4}@
  
  /* Check if my immediate successor is on the same socket */		@\linelabel{find-succ:5}@
  if (next->socket == mySocket) return next;					@\linelabel{find-succ:6}@

  cna_node_t *secHead = next;								@\linelabel{find-succ:7}@
  cna_node_t *secTail = next;								@\linelabel{find-succ:8}@
  cna_node_t *cur = next->next;								@\linelabel{find-succ:9}@

  /* Traverse the main queue */
  while (cur) {											@\linelabel{find-succ:11}@
    /* Check if cur is running on my socket */
    if (cur->socket == mySocket) {							@\linelabel{find-succ:13}@
      if (me->spin > 1) 										@\linelabel{find-succ:14}@
        ((cna_node_t *)(me->spin))->secTail->next = secHead;	        @\linelabel{find-succ:15}@
      else me->spin = (uintptr_t)secHead;						@\linelabel{find-succ:16}@
      secTail->next = NULL;									@\linelabel{find-succ:17}@
      ((cna_node_t *)(me->spin))->secTail = secTail;				@\linelabel{find-succ:18}@
      return cur;											@\linelabel{find-succ:19}@
    }
    secTail = cur;											@\linelabel{find-succ:21}@
    cur = cur->next;										@\linelabel{find-succ:22}@
  }
  
  return NULL;											@\linelabel{find-succ:23}@
}

/* Long-term fairness threshold */
#define THRESHOLD (0xffff)
int keep_lock_local() { return pseudo_rand() & THRESHOLD; }			@\linelabel{keep-lock-local:1}@

\end{lstlisting}
\caption{Auxiliary functions called from the Unlock procedure. }
\label{fig:unlock-aux}
\end{figure}

In this section, we provide the C-style pseudo-code for the CNA lock structures and implementation.
We assume sequential consistency for clarity.
Our actual implementation uses volatile keywords and memory fences as well as
padding (to avoid false sharing) where necessarily.

The CNA lock structure as well as the structure of the CNA node are provided in Figure~\ref{fig:structures}.
We note that the CNA queue node structure contains a few extra fields compared to MCS (\code{secondaryTail} and \code{socket}).
The space of queue node structures, however, is almost never a practical concern, since
those structures can be reused for different lock acquisitions, and between different locks.
Normally, each thread would have a preallocated (small) number of such structures in a thread-local storage.
In the Linux kernel, for example, each thread (or more precisely, each CPU) 
has exactly four such statically preallocated structures~\cite{Long13}.
Nevertheless, we discuss an optimization in~\secref{sec:opts} that eliminates the \code{socket} field, albeit
mostly for performance considerations.
Note that the CNA lock instance itself requires one word only --- a pointer to the tail of the main queue.

The details of the lock procedure are given in Figure~\ref{fig:lock}.
Those familiar with the MCS lock implementation will note a striking similarity between the lock procedure of CNA to the one of MCS.
The only differences are in \lineref{lock:8}, which makes sure the \texttt{spin} field is set (to $1$) before returning, and in 
\lineref{lock:3} and \lineref{lock:10}, which initialize and record the current socket number of a thread in its node structure.
Setting the \texttt{spin} field in \lineref{lock:8} is done so we pass a non-zero value to the successor in the unlock procedure (as explained below).
As for recording the socket number in \lineref{lock:10}, we note that on the x86 platform, one could use an efficient \code{rdtscp} instruction for that.
If a platform does not support a lightweight way to retrieve the current socket number, one could store this information in a thread local variable, and 
refresh it periodically, e.g., every 1K lock acquisitions.
Note that if a thread is migrated and its actual socket number is different from the one recorded in its node structure, this might have only 
performance implications but not correctness.
Finally, we note that recording the socket number takes place only if the thread finds another node in the (main) queue (cf.~\lineref{lock:8}).
In other words, when the lock is not contended, this line does not add any overhead to the Lock procedure. 

The pseudo-code for the unlock procedure is presented in Figure~\ref{fig:unlock} and auxiliary functions it uses are in Figure~\ref{fig:unlock-aux}.
Like the unlock procedure of the MCS lock, it starts by checking whether there are other nodes in the main queue (\lineref{unlock:3}).
If not, it attempts to set the \code{tail} of the main queue to either NULL (\lineref{unlock:8}) or the tail of the secondary queue if the latter is not empty (\lineref{unlock:12}).
Note that, similarly to MCS, an atomic compare-and-swap (CAS) instruction is needed to make sure no thread has enqueued itself into the main queue
between the if-statement in \lineref{unlock:3} and the update of the \code{tail} pointer.
If that does happen, the lock holder $h$ waits for its successor to update the \code{next} pointer in $h$'s node (\lineref{unlock:19}).

If the main queue is not empty, the lock holder $h$ determines the next lock holder by invoking the \code{find\_successor} function.
There, it traverses the main queue and searches for a thread running on the same socket as $h$.
If it finds one, it moves all the threads between $h$ and that successor to the secondary queue (\linerangeref{find-succ:14}{find-succ:18}).
Otherwise, it returns NULL without updating the secondary queue (\lineref{find-succ:23}).

If a successor on the same socket is found (and the \code{keep\_lock\_local} function returns a nonzero number), 
the lock is handed over to that successor (\lineref{unlock:23}).
(Note that if $h$ has entered an empty queue, \code{me->spin} will contain $1$ (cf.~\lineref{lock:8});
therefore, in \lineref{unlock:23} we always store a non-zero value into \code{succ->spin}).
Otherwise, it is handed to the first node in the secondary queue, 
after connecting the tail of that queue to the $h$'s successor in the main queue (\linerangeref{unlock:26}{unlock:27}).
(Note that it would be correct to set \code{succ->secTail} to NULL after \lineref{unlock:26}, but this is unnecessary 
since \code{succ} gets the lock and will not read this field. 
This explains why in \figref{fig:scheme}~(g) $t2$'s \code{secondaryTail} remains to point $t6$.)
If the secondary queue is empty, the lock is handed over to $h$'s successor in the main queue (\lineref{unlock:29}).

\section{Optimizations}
\seclabel{sec:opts}
The implementation of the CNA lock admits a few simple optimizations.
First, we can encode the socket of a thread in the \code{next} pointer of its 
predecessor in the queue (cf.~\lineref{lock:11}).
This would obviate the access to \code{socket} in \code{find\_successor} (cf.~\lineref{find-succ:6} and \lineref{find-succ:13}), 
and thus avoid a cache miss(es) on the critical path.
This encoding is straightforward on machines with a small (two/four) number of sockets, in which pointers are 
($4$-byte) word aligned. 
On bigger machines with more sockets, one can allocate queue node structures with a proper alignment to make space for the socket encoding.

Second, when the contention on the lock is light, the (small) overhead of moving threads in and out of the 
secondary queue might not justify the gain of keeping the lock local on the same socket.
As a result, the performance of the CNA lock might suffer in this case.
To avoid that, we can reduce the probability of moving threads to the secondary queue (or more precisely, of calling the 
\code{find\_successor} function) when the secondary queue is empty, and simply hand over the lock to the successor of the
lock holder in the main queue.
In other words, we can introduce the so-called \emph{shuffle reduction} optimization with the following lines of pseudo-code 
placed between \lineref{unlock:20} and \lineref{unlock:21} in \code{cna\_unlock} (cf.~Figure~\ref{fig:unlock}):

\begin{lstlisting}[style=nonumbers]
if (me->spin == 1 && (pseudo_rand() & THRESHOLD2)) {
      me->next->spin = 1;
      return;
}
\end{lstlisting}

In our experiments with the shuffle reduction optimization enabled, we set \texttt{THRESHOLD2} to \texttt{0xff}.

Finally, as already mentioned, the socket number can be cached in a thread-local variable and 
refreshed periodically.
Similarly, instead of drawing a pseudo-random number in every invocation of \code{keep\_lock\_local},
a thread can store the drawn number in a thread-local variable and decrement it with every lock handover.
Once the number reaches $0$, the thread would redraw a new number, and have \code{keep\_lock\_local} return zero.

While we experiment and report initial findings on the shuffle reduction optimization
in \secref{sec:evaluation},
the other optimizations are fairly straightforward engineering tweaks left for future work.

\section{Evaluation}
\seclabel{sec:evaluation}
For user-space benchmarks, we integrated CNA into LiTL~\cite{GLQ16}, an open-source project\footnote{https://github.com/multicore-locks/litl}
that provides an implementation of dozens of various locks, including the MCS 
lock as well as several state-of-the-art NUMA-aware locks.
All locks in LiTL, as well as our CNA lock, are implemented as dynamic libraries 
conforming to the pthread mutex lock API defined by the POSIX standard.
This allows interposing those locks with any software that uses the standard API (through the LD\_PRELOAD mechanism) 
without modifying or even recompiling the software code.
For kernel experiments, we integrated the CNA lock into the kernel version 4.20.0-rc4.
The resulting qspinlock uses the same fast path (based on the test-and-set lock; see \secref{sec:qspinlock}) and the CNA lock as the slow path.
In other words, we modified the slow path acquisition function (\code{queued\_spin\_lock\_slowpath} in \code{qspinlock.c}) to use CNA instead of MCS.

In user-space experiments, we compared the CNA lock to the MCS lock, as well as to several 
state-of-the-art NUMA-aware locks, namely C-BO-MCS, C-PTL-TKT and C-TKT-TKT locks, which
are the variants of Cohort locks~\cite{DMS15}, HMCS lock~\cite{CFM15} and HYSHMCS lock~\cite{KMK17}.
All NUMA-aware locks were configured with similar fairness settings, that is, 
keeping the lock local to a socket for a similar number of lock handovers before passing the lock to a thread running on another socket.
We note that in all our experiments, HMCS and HYSHMCS produced similar performance except where noted, 
while C-BO-MCS typically performed best among all Cohort lock variants.
Therefore, for presentation clarity purposes only, we omit the results of HYSHMCS 
and Cohort lock variants but C-BO-MCS in subsequent plots.

In the kernel, we compare the existing MCS-based qspinlock implementation to the new one based on CNA.
Note that we could not integrate any of the state-of-the-art NUMA-aware locks into the kernel as they require substantially 
more than $4$ bytes of space.

We run our experiments on a system with two Intel Xeon E5-2699 v3 sockets featuring 
18 hyperthreaded cores each (72 logical CPUs in total) and running Ubuntu 18.04.
To validate our conclusions beyond two sockets, we also run experiments on 
a system with four Intel Xeon E7-8895 v3 sockets featuring 144 logical CPUs in total 
and running Ubuntu 18.04.
For space considerations, we focus our presentation on the two-socket system and 
demonstrate only a few results from the four-socket one.
Thus, if not specified explicitly otherwise, the reported numbers were measured on the two-socket system.
We note, however, that in all experiments, the results on both systems were qualitatively similar.

In our experiments, we do not pin threads to cores, relying on the OS to make its choices. 
In all user-space experiments, we employ a scalable memory allocator~\cite{ADM11}.
In all kernel-space experiments, unless noted otherwise, we compile the kernel in the default configuration. 
In all experiments, we disable the turbo mode to avoid the effects that mode may have
(which varies with the number of threads) on the results. 
We vary the number of threads in each experiment from $1$ to $70$ ($142$) on the two-socket system 
(four-socket, respectively), leaving a few spare logical CPUs for any occasional 
kernel activity that might otherwise skew the results by preempting the lock holder thread.
Each reported experiment has been run $5$ times in exactly the same configuration. 
Presented results are the average of results reported by each of those $5$ runs.
The standard deviation of the measured numbers was less than $3\%$
from the average for the vast majority of the results, and always less than $10\%$ for all the results. 

Our evaluation is posed to answer the following questions: 
how the CNA lock compares to MCS and NUMA-aware locks in a user-space 
microbenchmark in terms of throughput and cache miss rates,
and how its long-term fairness fares to other locks (\secref{sec:user-space-microbenchmark}).
We also evaluate the impact of the shuffle reduction optimization discussed in 
\secref{sec:opts}.
Next, we evaluate the CNA lock with a few user-land benchmarks (\secref{sec:leveldb} and \secref{sec:kyoto}), 
and finally explore the impact it has on the Linux kernel performance through 
several kernel microbenchmarks (\secref{sec:kernel}).

\subsection{User-space benchmarks}
\subsubsection{Key-value map microbenchmark}
\seclabel{sec:user-space-microbenchmark}
We consider a simple key-value map implemented on top of an AVL tree protected with a single lock.
The benchmark includes operations for inserting, removing
and looking up keys (and associated values) stored in the key-value map (tree).
After initial warmup, not included in the measurement interval,
all threads start running at the same time, and apply  
operations chosen uniformly and at random
from the given operation mix, with keys chosen uniformly
and at random from the given range. At the end of the measured time
period (lasting 10 seconds), the total number of operations
is calculated, and the throughput is reported. 
The key-value map is pre-initialized to contain roughly half of the key range.
The benchmark allows varying the key range 
(effectively controlling the initial size of the key-value map)
as well as the amount of the external work, i.e., the
duration of a non-critical section (simulated by a pseudo-random
number calculation loop) between operations on the map.

\figref{fig:neelam-avl-tree-tput} shows the results of an experiment with the key range of $1024$ 
and an operation mix of $80\%$ lookups and $20\%$ updates (split evenly between
inserts and removes).
(We note that the size of the key range as well as the mix of operations mainly affects the length of the critical section;
we also experimented with different key range sizes and operation mixes, and the results were qualitatively the same.)
In this experiment, threads do not perform any external work, which results in 
substantial contention on the lock protecting the tree and absolutely no scalability.
The performance of the MCS lock drops abruptly between one and two threads, 
as the lock becomes contended with threads running on different sockets. 
This happens to all other locks but C-BO-MCS.
In the case of C-BO-MCS, the global (high-level) lock is backoff test-and-set, which in this case performs well 
since the same thread manages to acquire the lock repeatedly and effectively starve the other one.
We note that backoff-based locks are known to be unfair~\cite{RH03,LNS06}, with backoff timeouts hard to tune for optimal performance~\cite{JAS09, LA93}.

Beyond two threads, the performance of the MCS stays mostly flat.
At the same time, CNA lock matches the performance of MCS for $1$ and $2$ threads, and then improves afterwards 
as it starts taking advantage of its NUMA-awareness.
Notably, with more than $4$ threads CNA performs better that or on par with all variants of Cohort locks and 
only lags behind HMCS (and HYSHMCS) by a narrow margin.
Overall, it achieves $39\%$ speedup over MCS at $70$ threads.

\begin{figure}
\includegraphics[width=1\linewidth]{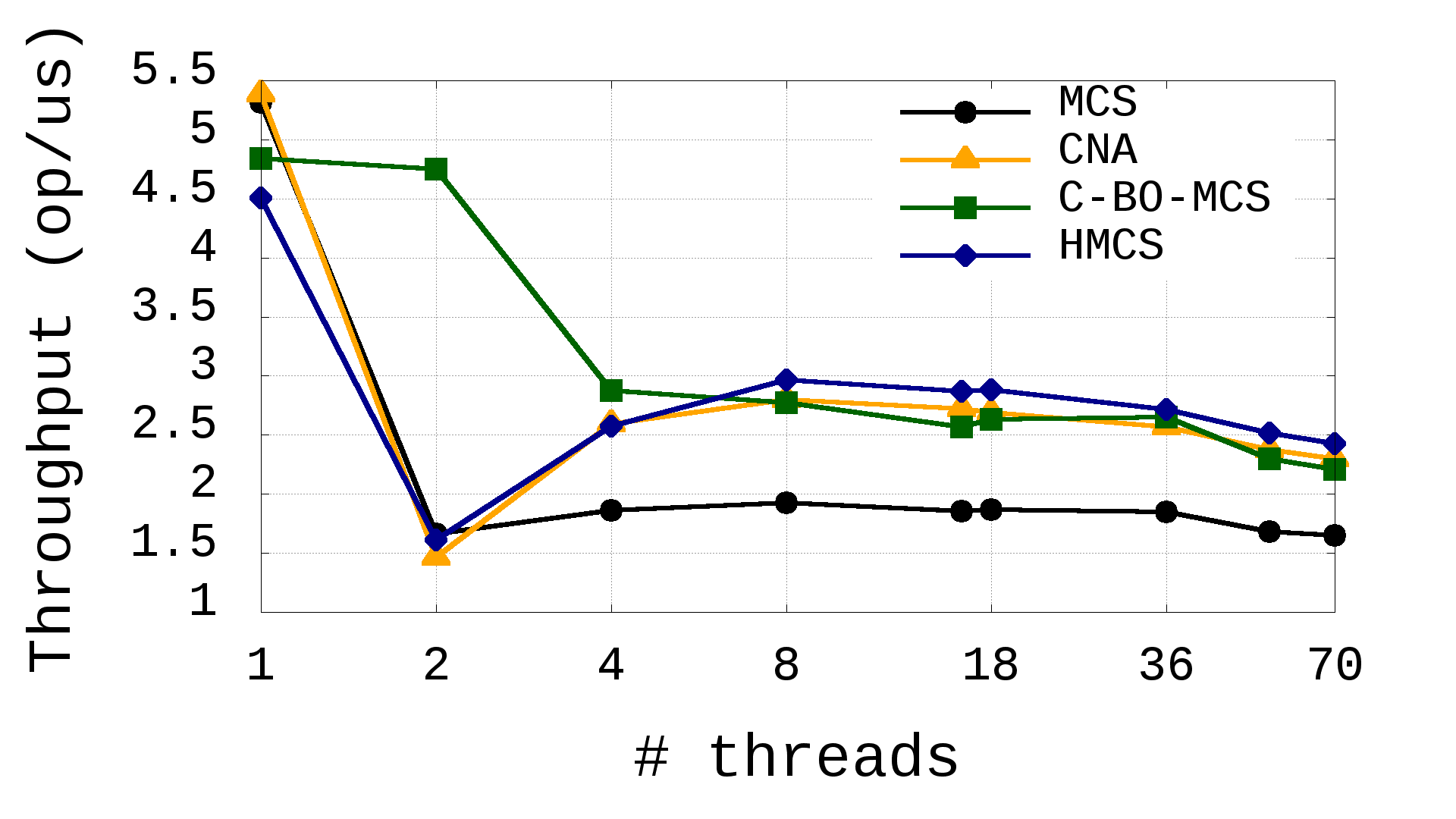}
\caption{Total throughput for the key-value map microbenchmark.}
\figlabel{fig:neelam-avl-tree-tput}
\end{figure}

We note that the workload in \figref{fig:neelam-avl-tree-tput} includes relatively few writes in the critical section.
We expect, however, that as the number of writes  into shared data increases, 
a NUMA-aware lock admission policy would yield even higher benefits, as
it would keep the shared data (in addition to the lock word itself) from migrating frequently between sockets. 
We confirmed that by experimenting with a workload composed of update-only operations.
While the results (not shown) generally follow the same pattern as \figref{fig:neelam-avl-tree-tput}, 
all NUMA-aware locks perform better relatively to the MCS lock.
In particular, CNA achieves the speedup of $50\%$ over MCS at $70$ threads.

\remove{
\figref{fig:neelam-avl-tree-tput}~(b) shows the results of the same experiment with an operation mix that does not include lookup operations.
As a result, threads performs more writes into shared data in their critical sections.
Therefore, we expect that a NUMA-aware lock admission policy would yield even higher benefits in this case, as
it would keep the shared data (in addition to the lock word itself) from migrating frequently between sockets.  
The results in \figref{fig:neelam-avl-tree-tput}~(b) confirm this hypothesis, and while generally following the same pattern as in \figref{fig:neelam-avl-tree-tput}~(a), 
all NUMA-aware locks perform better relatively to the MCS lock.
In particular, CNA achieves the speedup of $45\%$ over MCS at $70$ threads.
}

\begin{figure}
\includegraphics[width=1\linewidth]{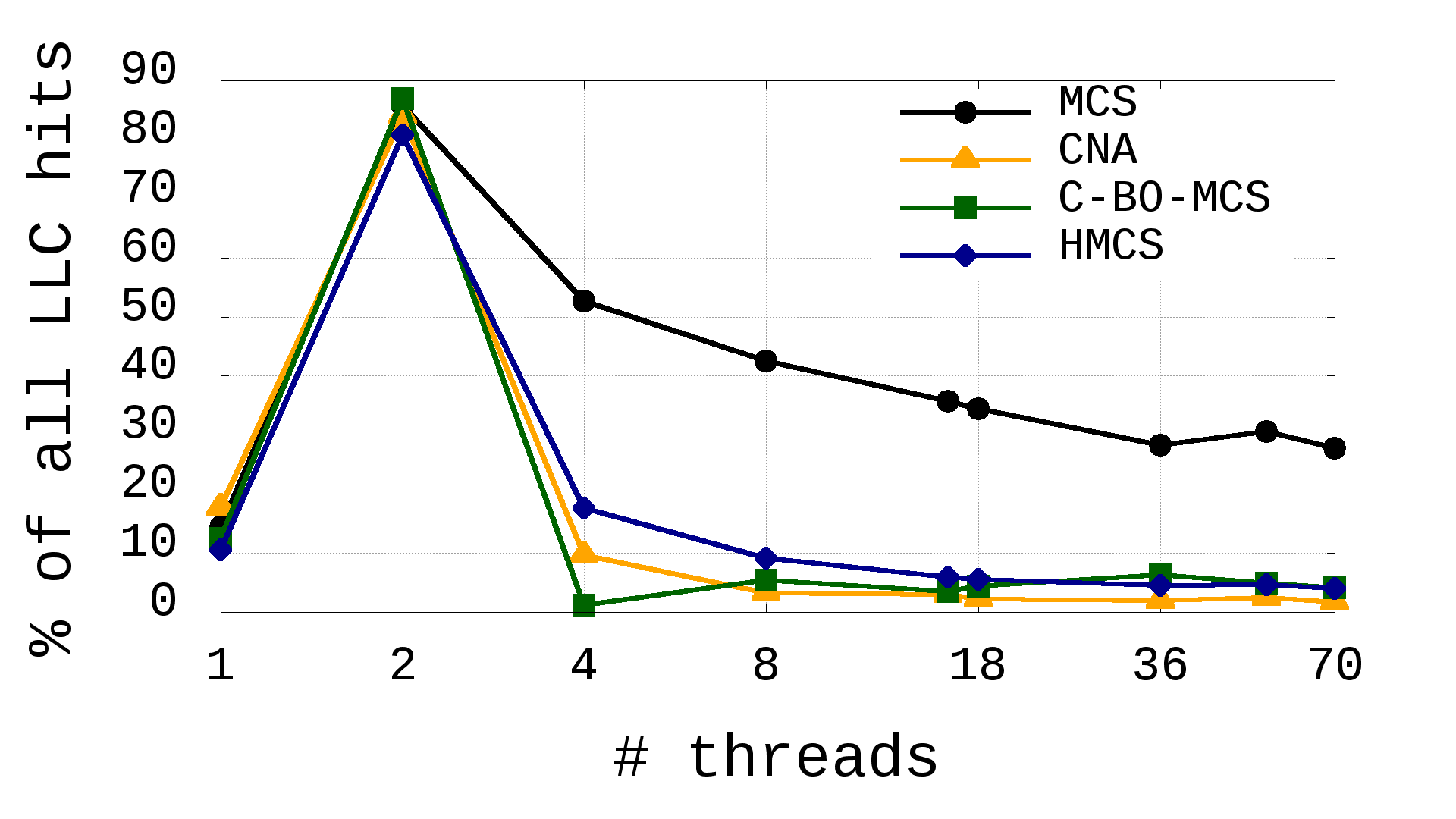}
\caption{LLC load miss rate for the key-value map microbenchmark.}
\figlabel{fig:neelam-avl-tree-cache-miss-rate}
\end{figure}

\begin{figure}
\includegraphics[width=1\linewidth]{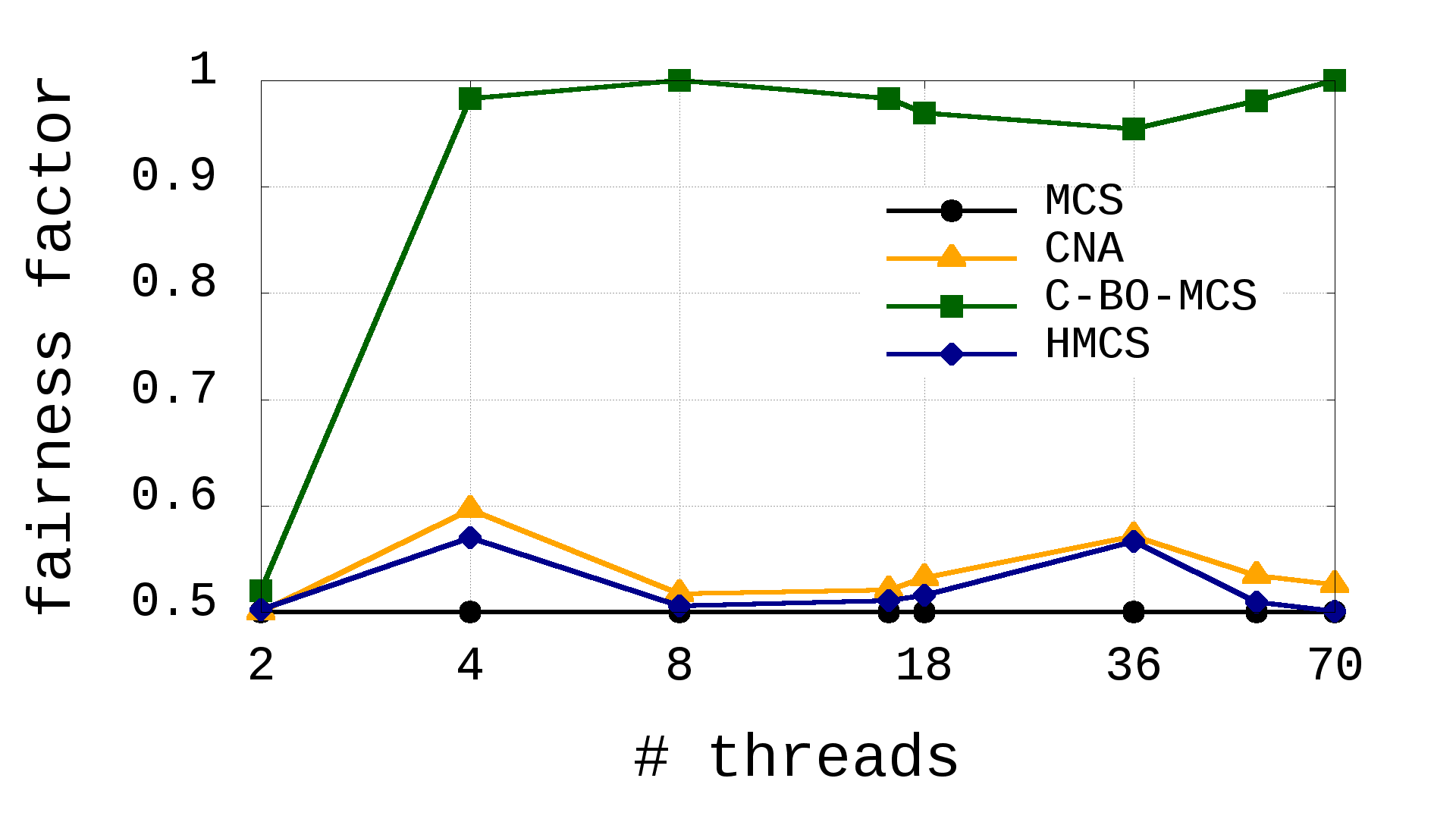}
\caption{Long-term fairness for the key-value map microbenchmark.}
\figlabel{fig:neelam-avl-tree-fairness}
\end{figure}

\figref{fig:neelam-avl-tree-cache-miss-rate} provides data on the LLC load miss rates (as measured by \code{perf}) for the evaluated locks.
(The presented data is for the workload with $20\%$ updates shown in \figref{fig:neelam-avl-tree-tput}.
Unless specified otherwise, the rest of the section considers this workload only, for brevity.)
The rates correlate with the throughput results in \figref{fig:neelam-avl-tree-tput}.
We note the sharp increase in LLC load miss rate between one and two threads (corresponding to the performance collapse at the same exact interval).
Afterwards, MCS exhibits a high LLC load miss rate unlike all other NUMA-aware locks, including CNA.

Naturally, one may ask how the long-term fairness of all the locks is affected by their NUMA-awareness (or the lack of it).
There are several ways to measure the fairness of a lock admission policy; 
in this work we present the data in the form of a \emph{fairness factor}.
To calculate this factor, we sort the number of operations performed by 
each thread as reported at the end of the experiment.
Then we divide the total number of the first half of the threads (in the sorted decreasing 
order of their number of operations) by the total number of operations.
Thus, the resulting fairness factor is a number in the range of $0.5$ and $1$, with a strictly 
fair lock yielding a factor of $0.5$ and a strictly unfair lock yielding a factor close to $1$.

\figref{fig:neelam-avl-tree-fairness} shows the fairness factor for each of the locks.
As expected, MCS with its strict FIFO admission policy maintains the fairness factor of $0.5$ across all thread counts.
HMCS has a fairness factor close to that, while C-BO-MCS has the factor close to $1$.
The latter is an example of the starvation behavior by a backoff test-and-set lock mentioned above.
The CNA lock achieves slightly higher fairness factor than MCS, yet those rates for the most part are well below $60\%$, suggesting that it preserves
the long-term fairness.
We note that like other state-of-the-art NUMA-aware locks, 
the CNA lock provides a knob to tune the fairness-vs-throughput tradeoff 
(cf.~\code{keep\_lock\_local} in Figure~\ref{fig:unlock-aux}).

\begin{figure}
\includegraphics[width=1\linewidth]{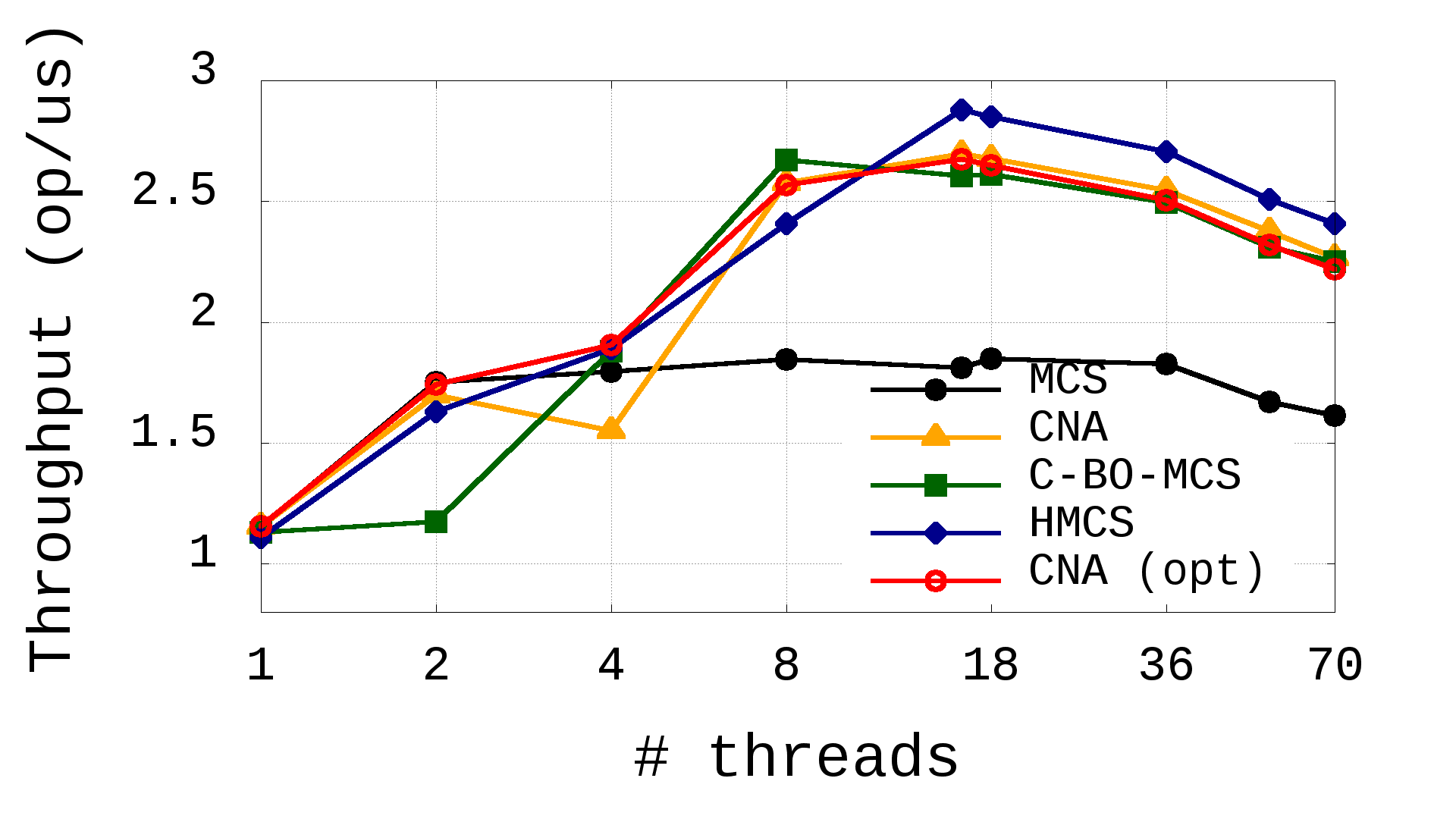}
\caption{Total throughput for the key-value map microbenchmark with non-critical work.}
\figlabel{fig:neelam-avl-tree-tput-with-extended-work}
\end{figure}

\begin{figure}
\includegraphics[width=1\linewidth]{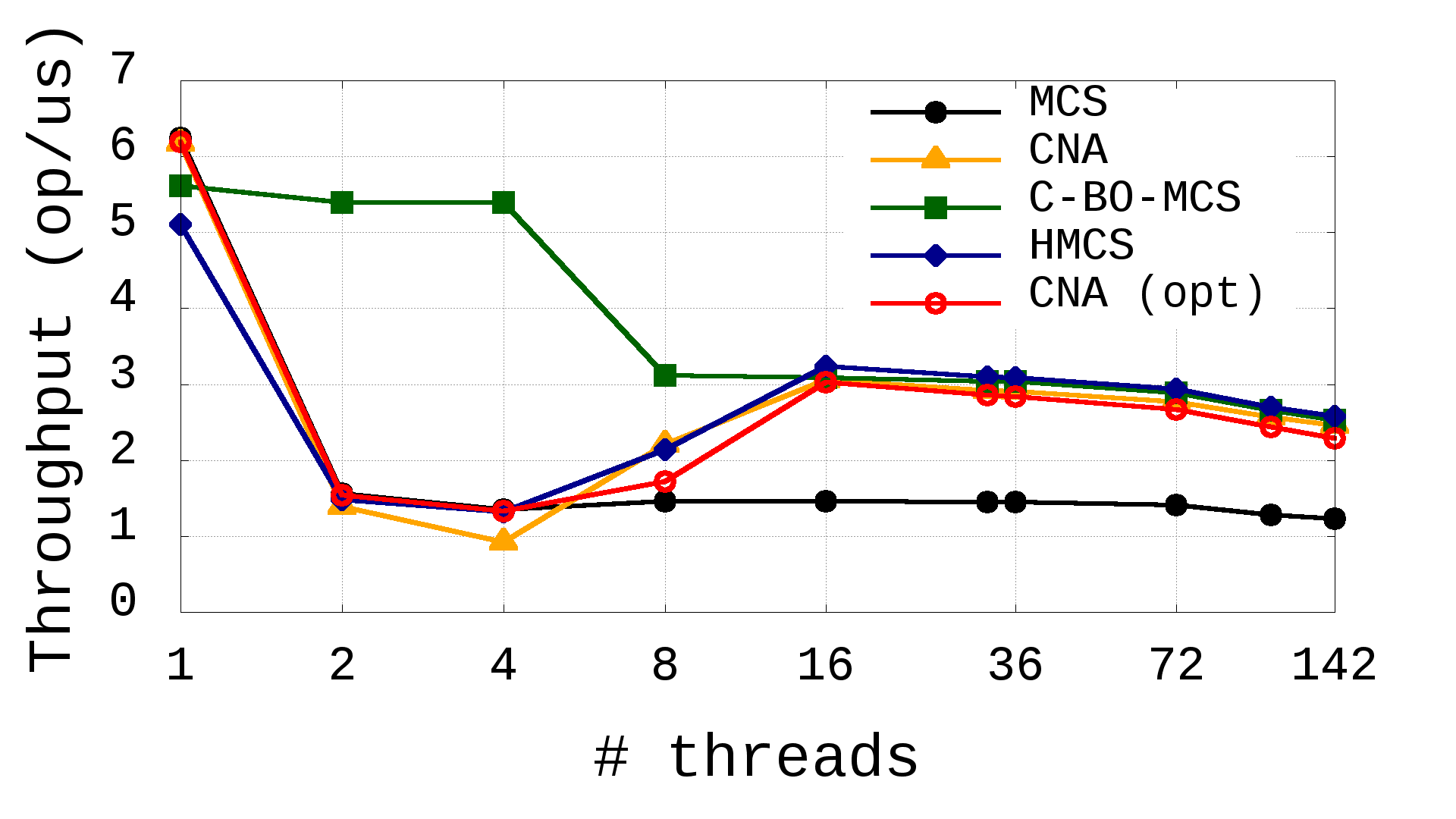}
\caption{Total throughput for the key-value map microbenchmark on a 4-socket machine (same workload as in \figref{fig:neelam-avl-tree-tput}).}
\figlabel{fig:x5-4-avl-tree-tput}
\end{figure}

\begin{figure*}
\subfloat[][Pre-filled DB]{\includegraphics[width=0.5\linewidth]{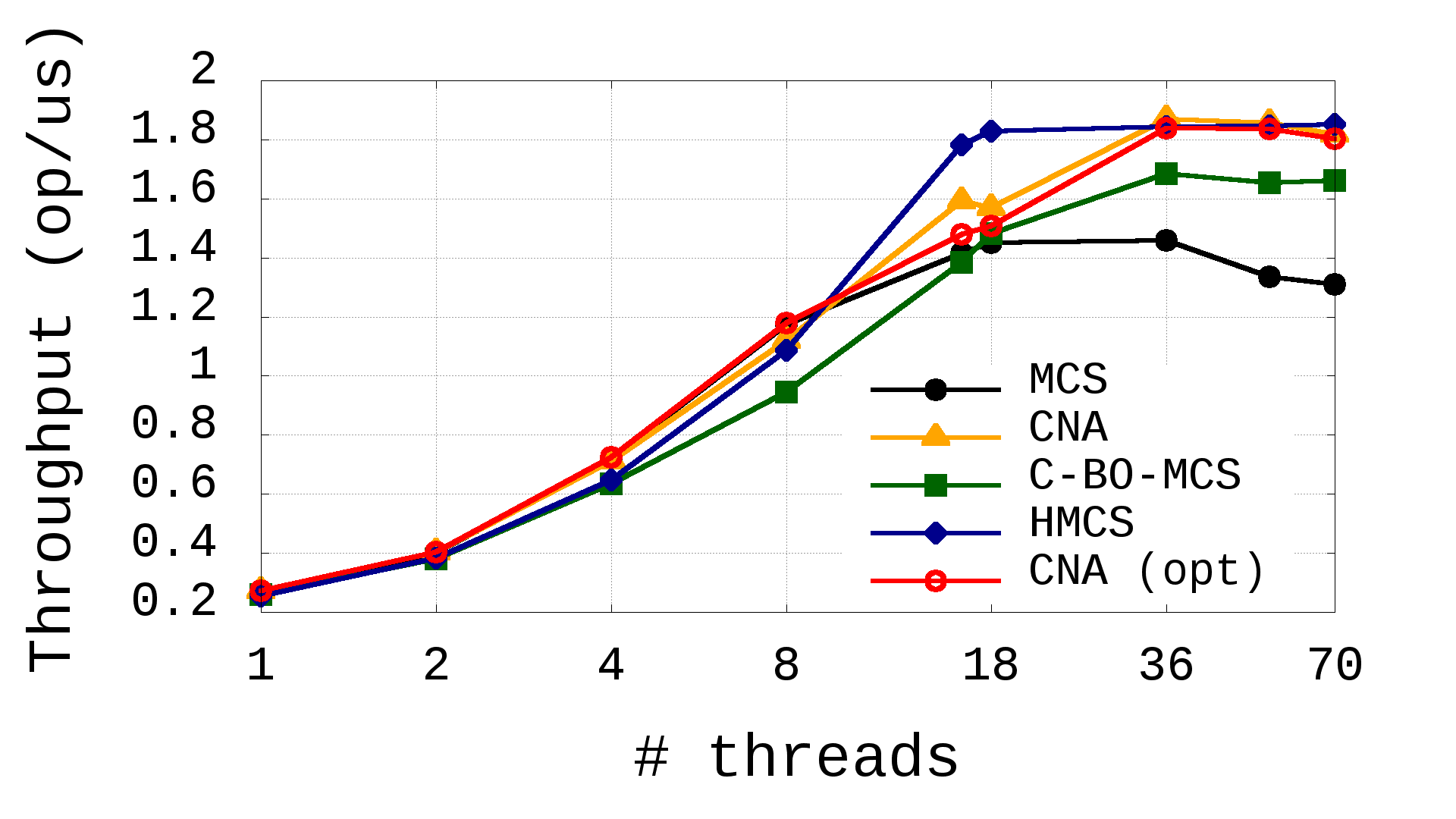}}
\subfloat[][Empty DB]{\includegraphics[width=0.5\linewidth]{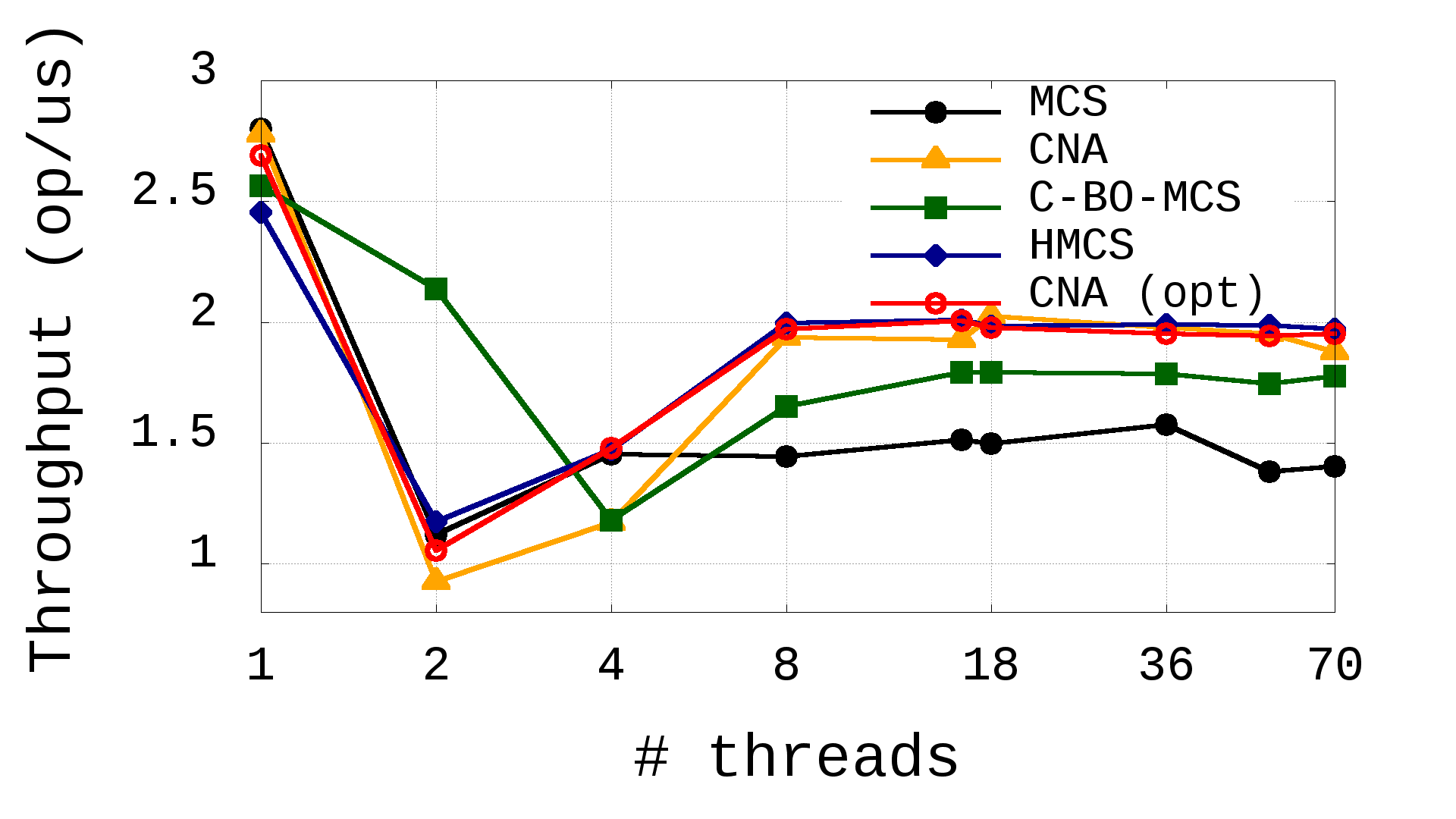}}
\caption{Total throughput for the leveldb benchmark.}
\figlabel{fig:neelam-levelDB}
\end{figure*}

In the next experiment, we add some amount of external (non-critical section) work 
that would reduce the lock contention 
and allow the benchmark to scale up to a small number of threads.
The results are presented in \figref{fig:neelam-avl-tree-tput-with-extended-work}.
Indeed, the performance of the benchmark when using the MCS lock increases between $1$ and $2$ threads, 
and then stays mostly flat.
All NUMA-aware locks scale up to 8-16 threads due to reduced cache miss rate and coherence traffic, 
and then maintain a substantial margin with respect to MCS.
Notably, CNA performs slightly worse (by $14\%$) than MCS at $4$ threads.
This is because at this thread count the contention is enough to occasionally shuffle waiting threads into the secondary queue, 
but not high enough to keep them there.
Intuitively, it results in paying continuously the (modest) price of accessing (and recording) socket information 
and restructuring the waiting queue without reducing remote cache misses in return.
We note that the LLC load miss rate for CNA (not shown for this experiment) matches the one for MCS at $4$ threads (but drops dramatically after that).
As the number of threads increases beyond $4$, CNA reaches and maintains a speedup of about $40\%$ over MCS, 
with a performance level between C-BO-MCS and HMCS.

\remove{
In the context of this experiment, we explore the potential of the \code{socket} embedding optimization described in \secref{sec:opts}.
We implement a variant of CNA in which the socket number is embedded in the \code{next} pointer of a queue node.
This variant is denoted as CNA (opt) in \figref{fig:neelam-avl-tree-tput-with-extended-work}.
The latter demonstrates that the \code{socket} embedding is not only useful to reduce (halve) the slowdown at $4$ threads with respect to MCS, 
but also to achieve better performance under contention, matching the performance level of HMCS. 
}

In the context of this experiment, we explore the potential of the shuffle reduction optimization described in \secref{sec:opts}.
We implement a variant of CNA in which, if the secondary queue is empty, the lock is handed over to the immediate successor in the main queue 
(and without searching for another thread running on the same socket) with high probability.
This variant is denoted as CNA (opt) in \figref{fig:neelam-avl-tree-tput-with-extended-work}.
With this optimization in place, CNA (opt) closes the gap from (and in fact, outperforms) MCS at $4$ threads, and matches the performance of CNA 
(without the optimization) at all other thread counts.
The superior performance over MCS at $4$ threads is the result of the shuffling that does take place once in a while, organizing threads' arrivals to the lock 
in a way that reduces the inter-socket lock migration without the need to continuously modify the main queue.
This is confirmed by LLC load miss rates, which are lower for CNA (opt) compared to MCS (and CNA) at $4$ threads.
We also collected statistics on how many times the main waiting queue is altered in CNA, and confirmed that the shuffle reduction optimization
indeed reduces this number by almost a factor of ten at $4$ threads (and has no impact at other thread counts).
We note, however, that the extent to which this occasional reorganization of waiting threads would have a positive effect on performance 
is expected to be application-dependent.

To validate our conclusions beyond two sockets, we also run experiments on 
a system with four Intel Xeon E7-8895 v3 sockets featuring 144 logical CPUs in total.
\figref{fig:x5-4-avl-tree-tput} shows the results of the experiment with the same workload as the one shown 
in \figref{fig:neelam-avl-tree-tput}.
While qualitatively all evaluated locks exhibited the same behavior, quantitatively the performance of CNA (and other NUMA-aware locks)
under contention improved relatively to MCS.
As an example, at $142$ threads CNA performs better than MCS by $97\%$.
We believe this is because the cost of a remote cache miss 
(fetching data from the LLC on another socket)
on this machine is higher than on the two-socket machine.
This can be seen by the drop in performance between $1$ and $2$ threads --- on the two-socket machine
the throughput of the MCS lock goes down from from $5.3$ ops/us to $1.7$ ops/us (cf.~\figref{fig:neelam-avl-tree-tput}), 
while on the four-socket machine it goes down from $6.2$ ops/us to $1.5$ ops/us (cf.~\figref{fig:x5-4-avl-tree-tput}).

\subsubsection{leveldb}
\seclabel{sec:leveldb}

In the next two sections we explore how the microbenchmark results discussed in \secref{sec:user-space-microbenchmark}
extend to real applications.
We start with leveldb, an open-source key-value storage library developed by Google.\footnote{https://github.com/google/leveldb}
Our experiments were done with the latest release (1.20) of the library, which also includes a built-in benchmark (\code{db\_bench}).

We used \code{db\_bench} to create a database with the default 1M key-value pairs. 
This database was used subsequently in the \code{readrandom} mode of \code{db\_bench}.
As its name suggests, the \code{readrandom} mode is composed of 
Get operations on the database with random keys. 
Each Get operation acquires a global database lock in order to take a consistent snapshot 
of pointers to internal database structures (and increment reference counters
to prevent the deletion of those structures while Get is running).
The search operation itself, however, executes without holding
the database lock, but acquires locks protecting (sharded)
LRU cache as it seeks to update the cache structure with the accessed key.
While the central database lock and internal LRUCache locks are known to be contended~\cite{Dice17},
the contention is spread over multiple locks.
 
We modified the \code{readrandom} mode to run for a fixed time (rather than run a certain number
of operations, so we could better control the runtime of each experiment). 
The reported numbers are the aggregated throughput for runs of $30$ seconds in the \code{readrandom} mode. 
\figref{fig:neelam-levelDB}~(a) presents the results, which show that  
as long as the the benchmark scales, all locks yield a similar performance 
(with only C-BO-MCS lagging slightly behind at $8$ threads).
However, once the scaling slows down, CNA outperforms MCS (and C-BO-MCS).
Overall, the performance pattern is somewhat similar to the one in \figref{fig:neelam-avl-tree-tput-with-extended-work}.
At the largest thread count, CNA outperforms MCS by $39\%$.

We also explore how increased contention on the database lock 
affects the speedup achieved by CNA.
To that end, we run the same \code{readrandom} mode with an empty database.
In this case, the work outside of the critical sections (searching for a key) is minimal and does not involve
acquiring any LRU cache lock.
The results are shown in \figref{fig:neelam-levelDB}~(b), and in general, are similar to the microbenchmark results 
with no external work in \figref{fig:neelam-avl-tree-tput}.
Once again, the shuffle reduction optimization proves useful at low thread counts.

\subsubsection{Kyoto Cabinet}
\seclabel{sec:kyoto}
We detail the results of the experiment with Kyoto Cabinet, 
another open-source key-value storage engine.\footnote{http://fallabs.com/kyotocabinet}
We use its built-in \code{kccachetest} benchmark run in a \code{wicked} mode, which exercises
an in-memory database with a random mix of operations.
Similarly to the approach taken by Dice~\cite{Dice17}, we modified the benchmark to use the standard POSIX pthread mutex
locks, which we interpose with evaluated locks from the LiTL library.
Similarly to leveldb, we modified the benchmark to run for a fixed
time and report the aggregated completed work. 
We also note that, originally, the benchmark sets the key range 
dependent on the number of threads, which makes the performance comparison across different thread counts challenging.
Therefore, we fixed the key range at a constant (10M) elements.  
Note that all those changes were also applied by Dice~\cite{Dice17}. 
The length of each run was $60$ seconds.

The results are presented in~\figref{fig:neelam-kyoto}.
The performance of \code{kccachetest} benchmark does not scale, and in fact, becomes worse as the contention grows.
Therefore, the best performance is achieved with a single thread, and CNA is the only lock to match the performance of MCS at that thread count.
As the number of threads grows beyond $4$, CNA (and other NUMA-aware locks) exploit the increasing lock contention, and outperform MCS.
At $36$--$70$ threads, CNA performs better than MCS by $28$--$43\%$.

\begin{figure}
\centering
\includegraphics[width=1\linewidth]{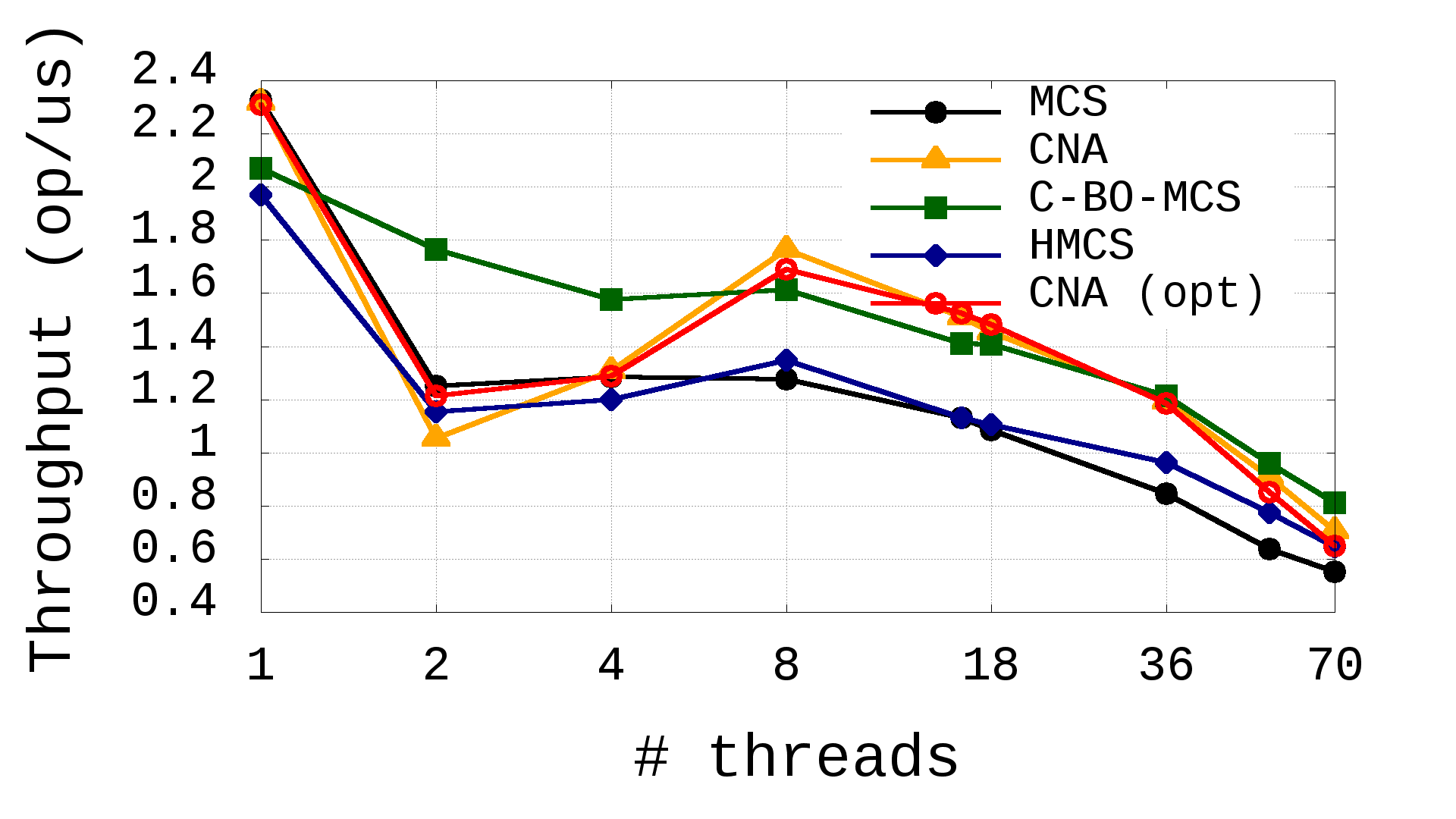}
\caption{Total throughput for Kyoto Cabinet.}
\figlabel{fig:neelam-kyoto}
\end{figure}

\textbf{In summary}, the user-space experiments show that CNA matches the performance of MCS with a single thread.
At higher contention, CNA beats MCS by a substantial margin and performs similarly to other state-of-the-art NUMA-aware locks,
while requiring far less space compared to them.

\subsection{Linux kernel}
\seclabel{sec:kernel}

\subsubsection{locktortture}
The \code{locktorture} benchmark is distributed as a part of the Linux kernel.\footnote{https://www.kernel.org/doc/Documentation/locking/locktorture.txt}
It is implemented as a loadable kernel module, and according to its documentation, 
``runs torture tests on core kernel locking primitives'', 
including \code{qspinlock}, the kernel spin lock.
It creates a given number of threads that repeatedly acquire and release the lock, with occasional short 
delays (citing the comment in the source code, ``to emulate likely code'') and occasional long 
delays (``to force massive contention'') inside the critical section.
At the end of a run (lasting $30$ seconds in our case), a total number of lock operations performed by all threads is reported.

\begin{figure*}
\centering
\subfloat[][lockstat disabled (default)]{\includegraphics[width=0.5\linewidth]{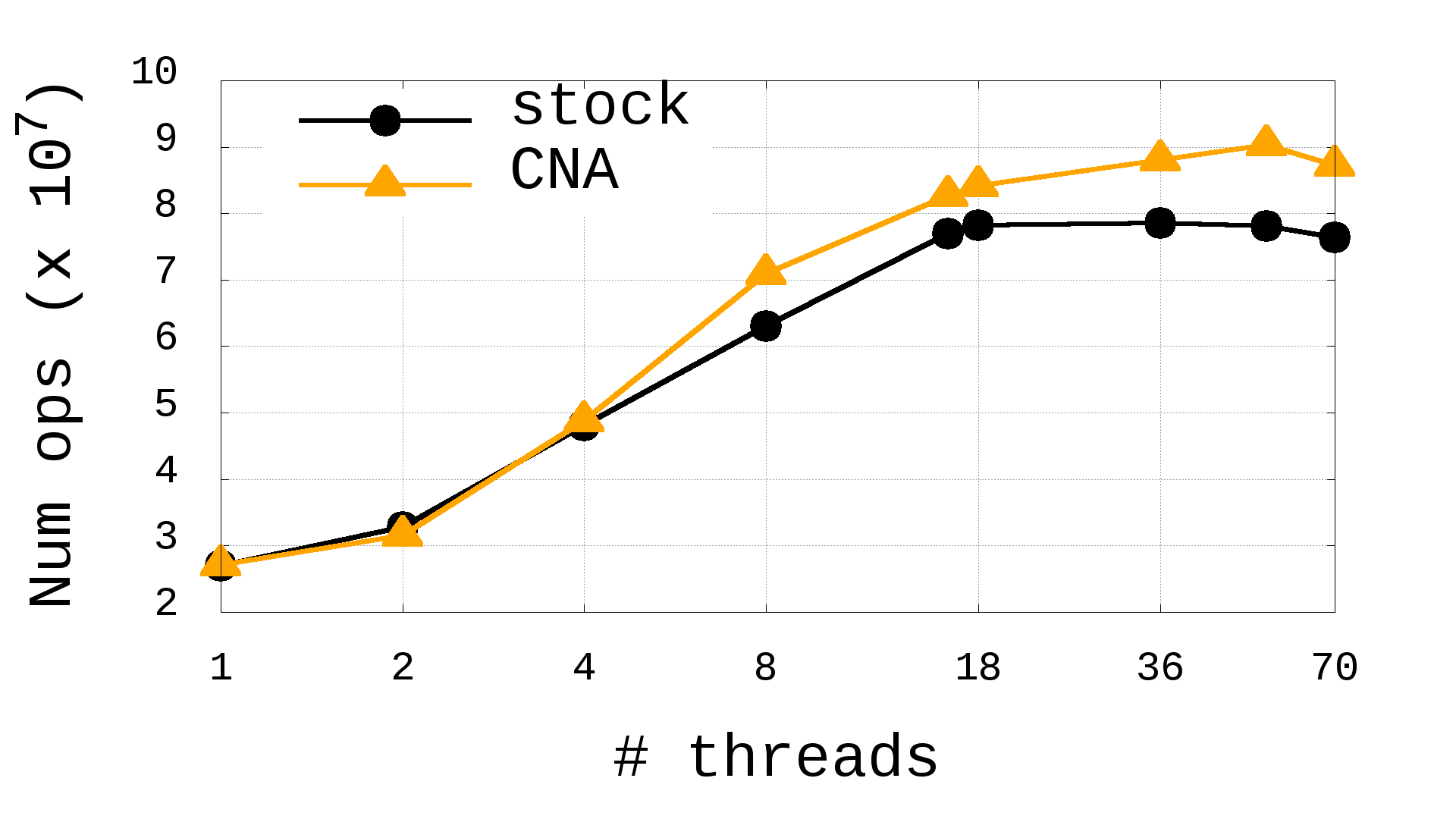}}
\subfloat[][lockstat enabled]{\includegraphics[width=0.5\linewidth]{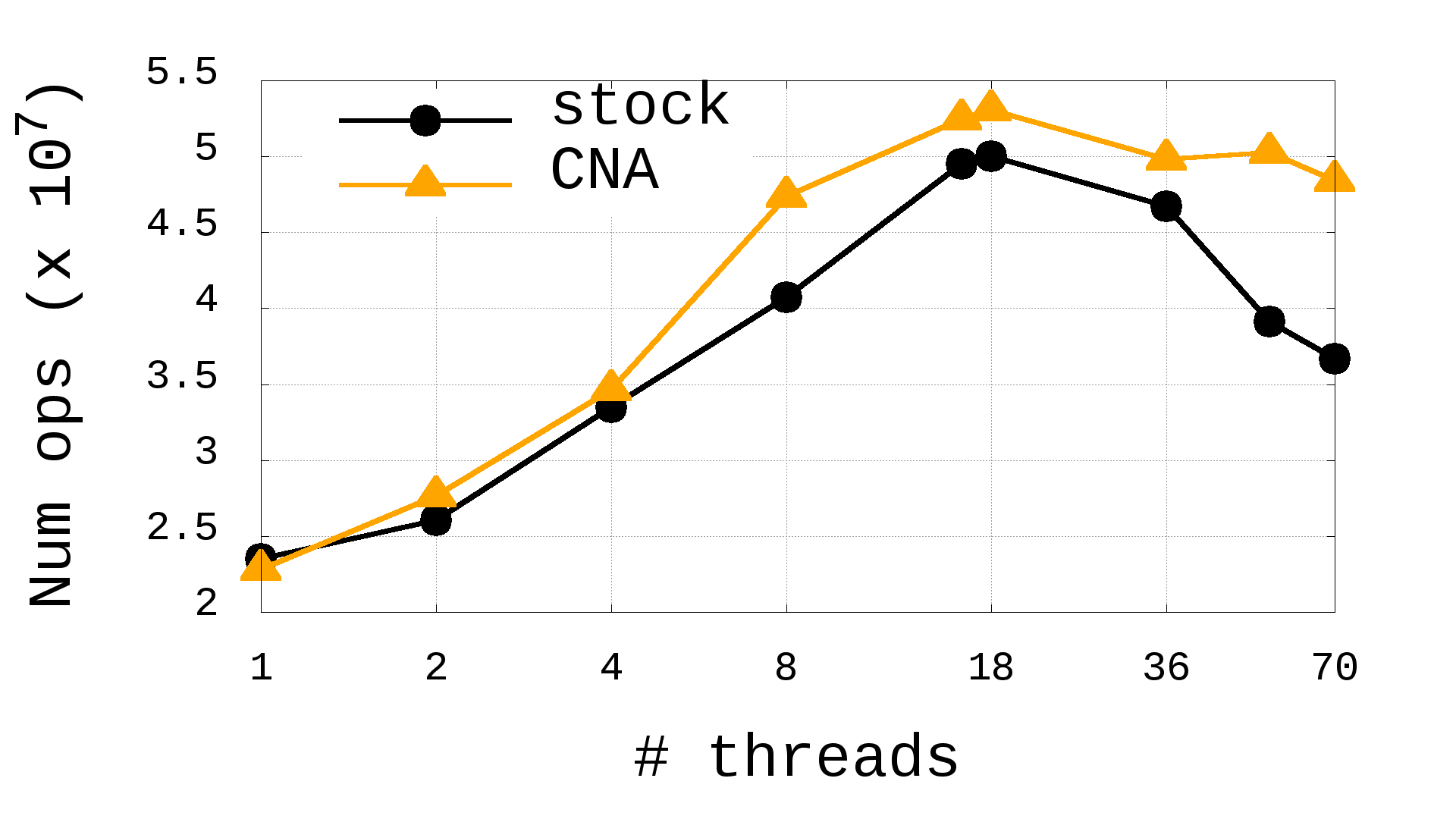}}
\caption{Performance results for \code{locktorture} on the 2-socket machine.}
\figlabel{fig:locktorture-neelam}
\end{figure*}

\begin{figure*}
\centering
\subfloat[][lockstat disabled (default)]{\includegraphics[width=0.5\linewidth]{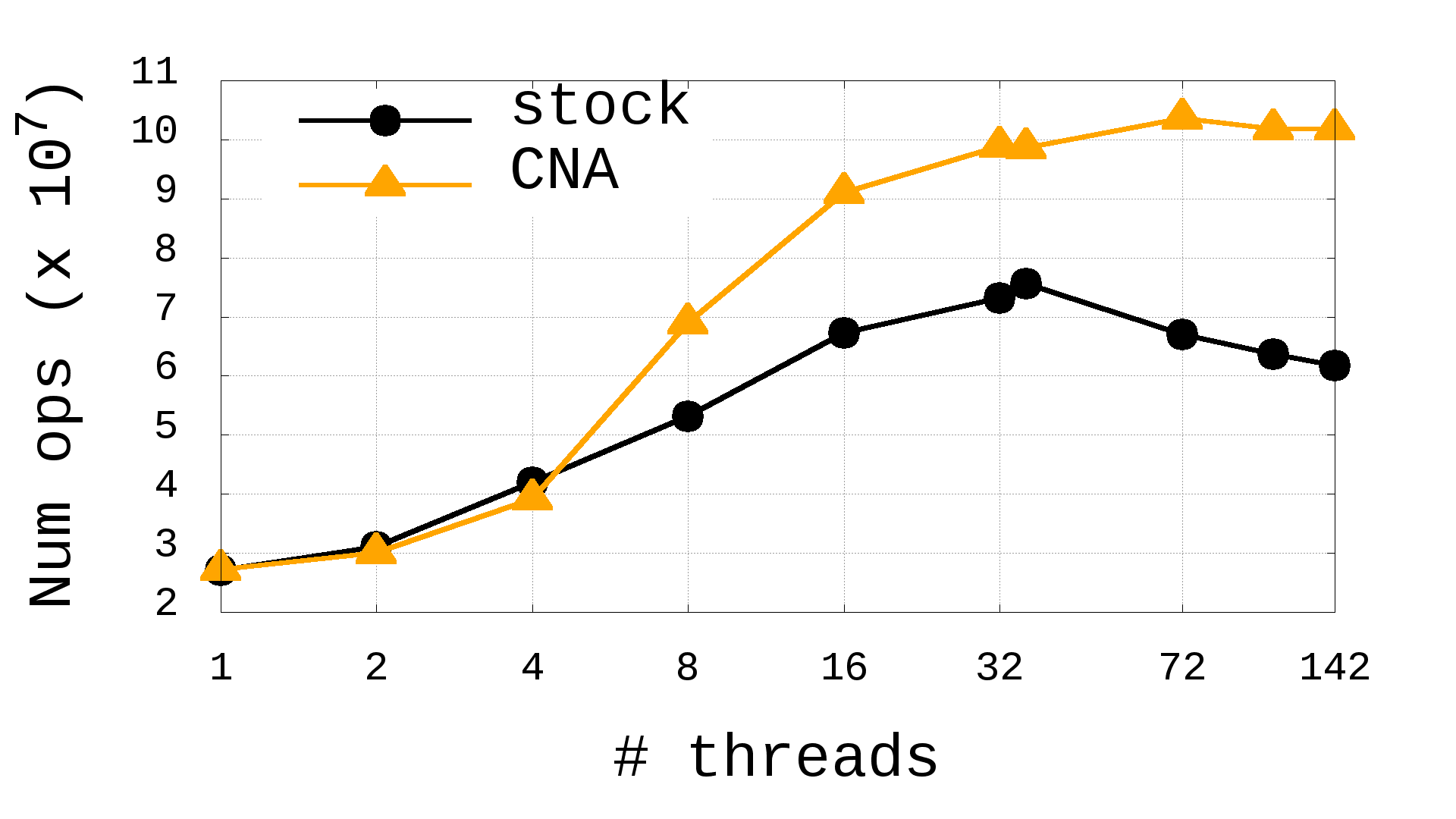}}
\subfloat[][lockstat enabled]{\includegraphics[width=0.5\linewidth]{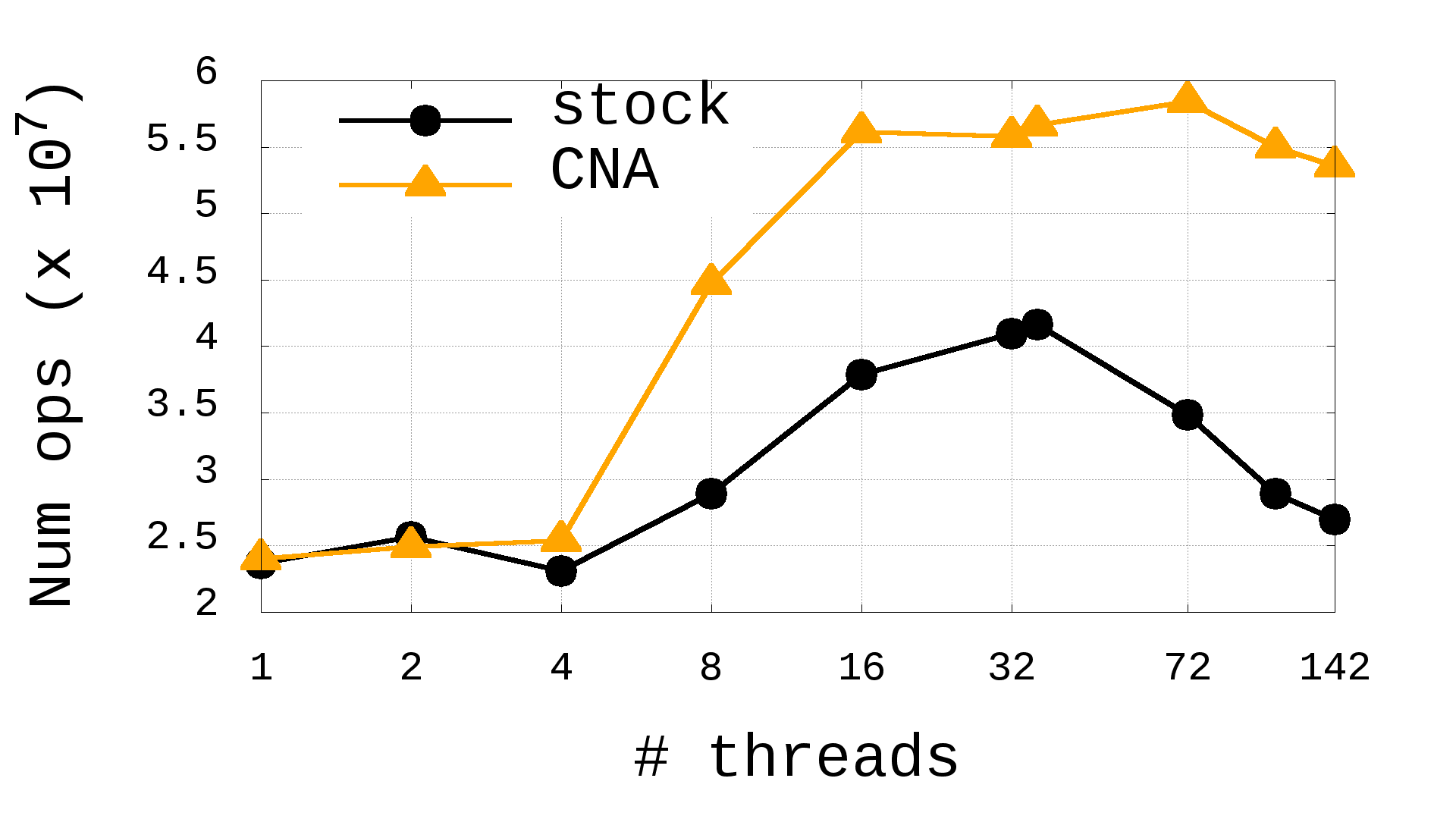}}
\caption{Performance results for \code{locktorture} on the 4-socket machine.}
\figlabel{fig:locktorture-ol-bur-x5-4}
\end{figure*}

The performance results for the \code{locktorture} benchmark are shown in \figref{fig:locktorture-neelam}~(a).
The unmodified Linux kernel is denoted as \code{stock}, while the version with CNA integrated into the \code{qspinlock} slow path
is denoted simply as \code{CNA}. 
Overall, \code{CNA} outperforms the \code{stock} version beyond $4$ threads, gaining $14\%$ at $70$ threads.

One of the benefits of a NUMA-aware lock is that it not only keeps the lock word local, but also it allows the shared 
data accessed by threads in their critical sections to stay on the same socket.
In \code{locktorture}, however, threads hardly access any shared data, as the critical section is emulated with
a random delay.
To demonstrate the impact of avoiding remote cache misses on shared data accesses without changing the 
 \code{locktorture} benchmark, we compiled the kernel with \code{lockstat}\footnote{\url{https://www.kernel.org/doc/Documentation/locking/lockstat.txt}} enabled. 
\code{Lockstat} is an optional built-in performance data collection mechanism used for debugging kernel locks performance.
After each lock acquisition, \code{lockstat} updates several shared variables, e.g., 
to keep track of the last CPU on which a given lock instance was acquired.
Those updates produce accesses to the shared data in the critical section and arguably represent more accurately
critical sections of real applications.
The performance results for the \code{locktorture} benchmark on the kernel compiled with \code{lockstat} enabled 
are shown in \figref{fig:locktorture-neelam}~(b).
The gap between \code{CNA} and \code{stock} grows up to $32\%$ at $70$ threads.

\figref{fig:locktorture-ol-bur-x5-4} shows the results of the \code{locktorture} benchmark on the four-socket machine,
in the default configuration and with \code{lockstat} enabled, respectively.
The results are similar to those measured on the two-socket machine, albeit the gap between 
\code{CNA} and \code{stock} is larger, likely due to a larger cost of a remote cache miss on the four-socket machine.
In particular, in the default configuration \code{CNA} outperforms \code{stock} by up to $65\%$ (\figref{fig:locktorture-ol-bur-x5-4}~(a)), 
while in the configuration with \code{lockstat} enabled the gap grows up to $99\%$ (\figref{fig:locktorture-ol-bur-x5-4}~(b)) at $142$ threads.

\begin{figure*}
\subfloat[][lock1\_threads]{\includegraphics[width=0.5\linewidth]{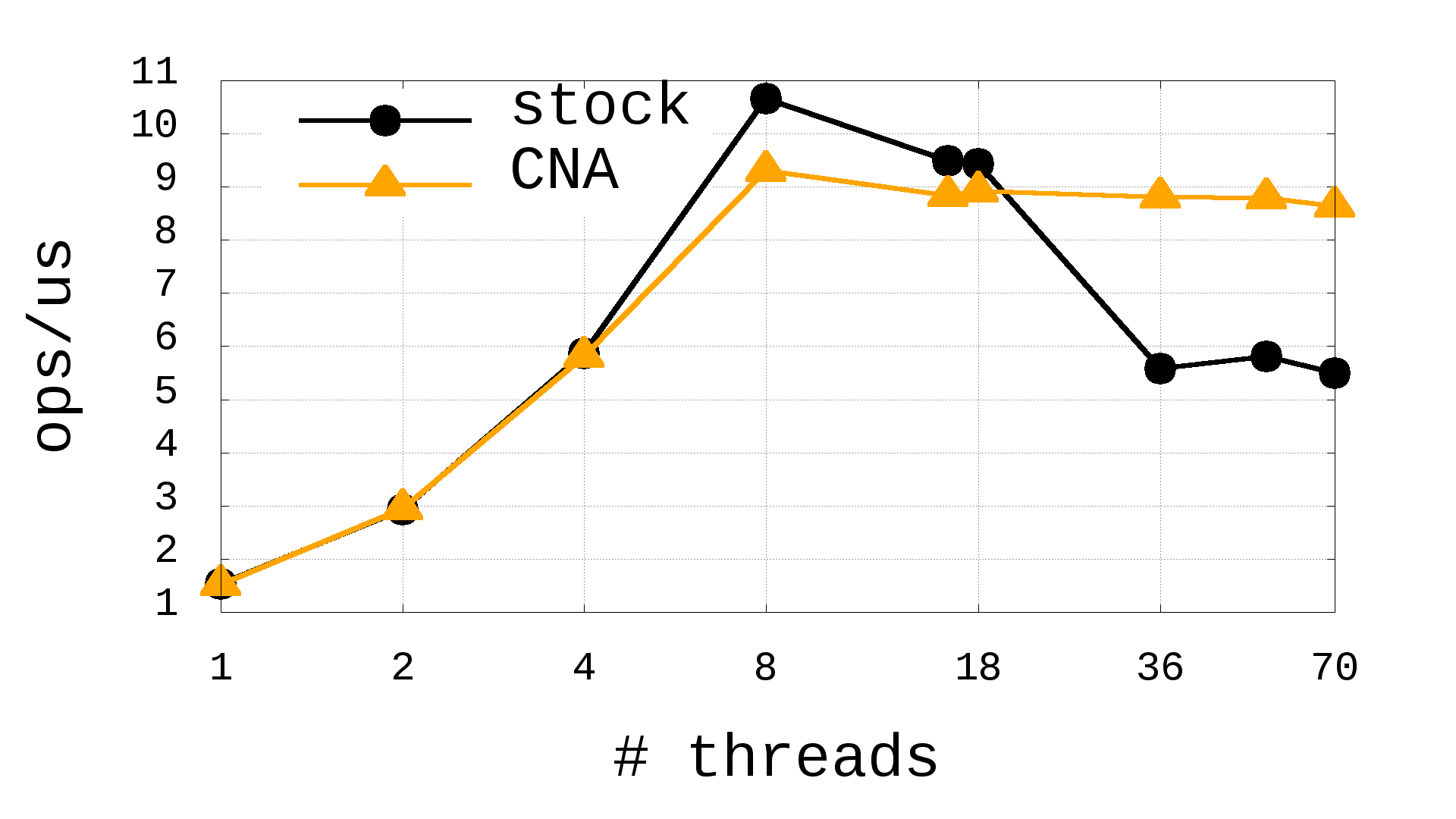}}
\subfloat[][lock2\_threads]{\includegraphics[width=0.5\linewidth]{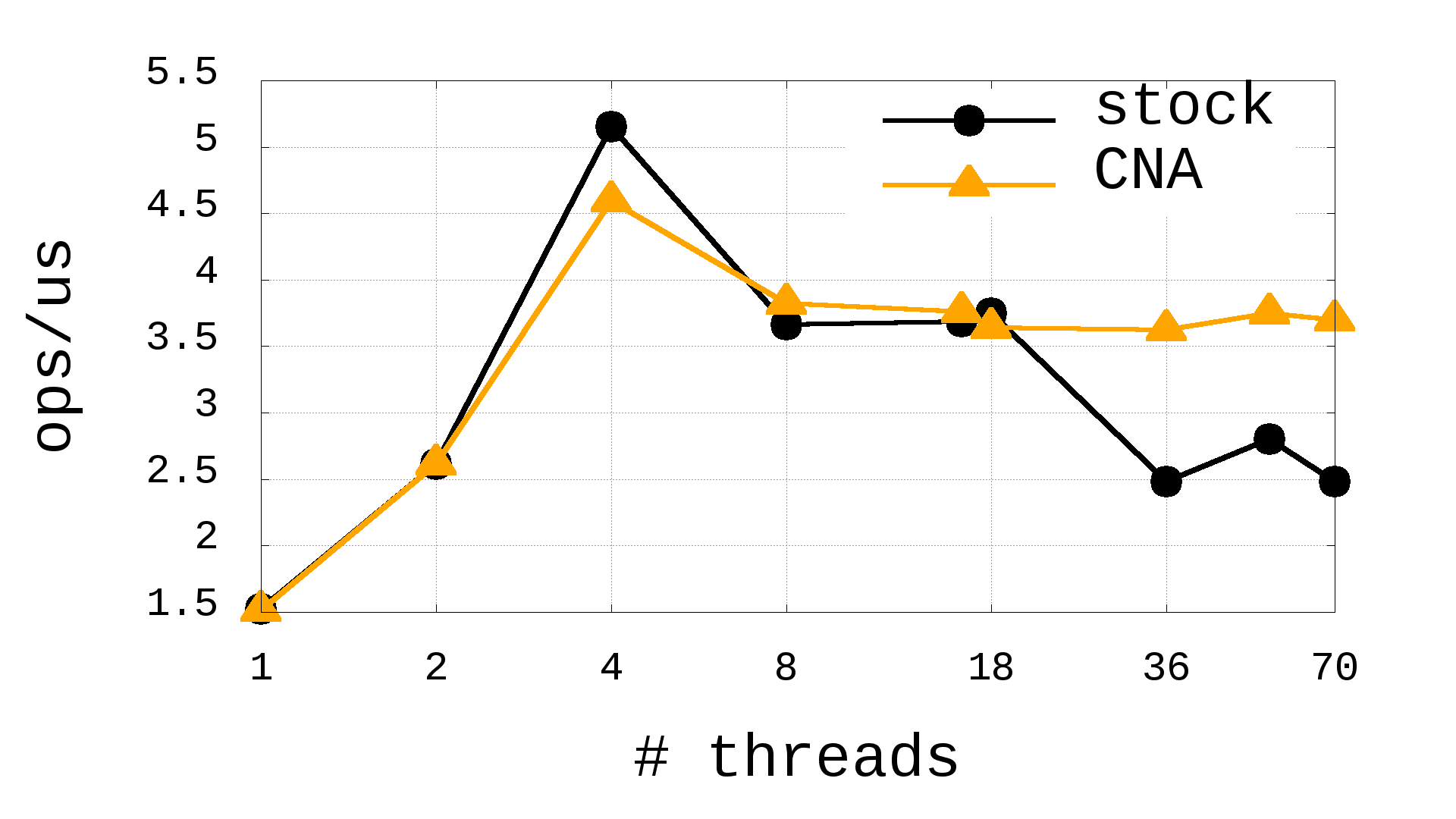}}\\
\subfloat[][open1\_threads]{\includegraphics[width=0.5\linewidth]{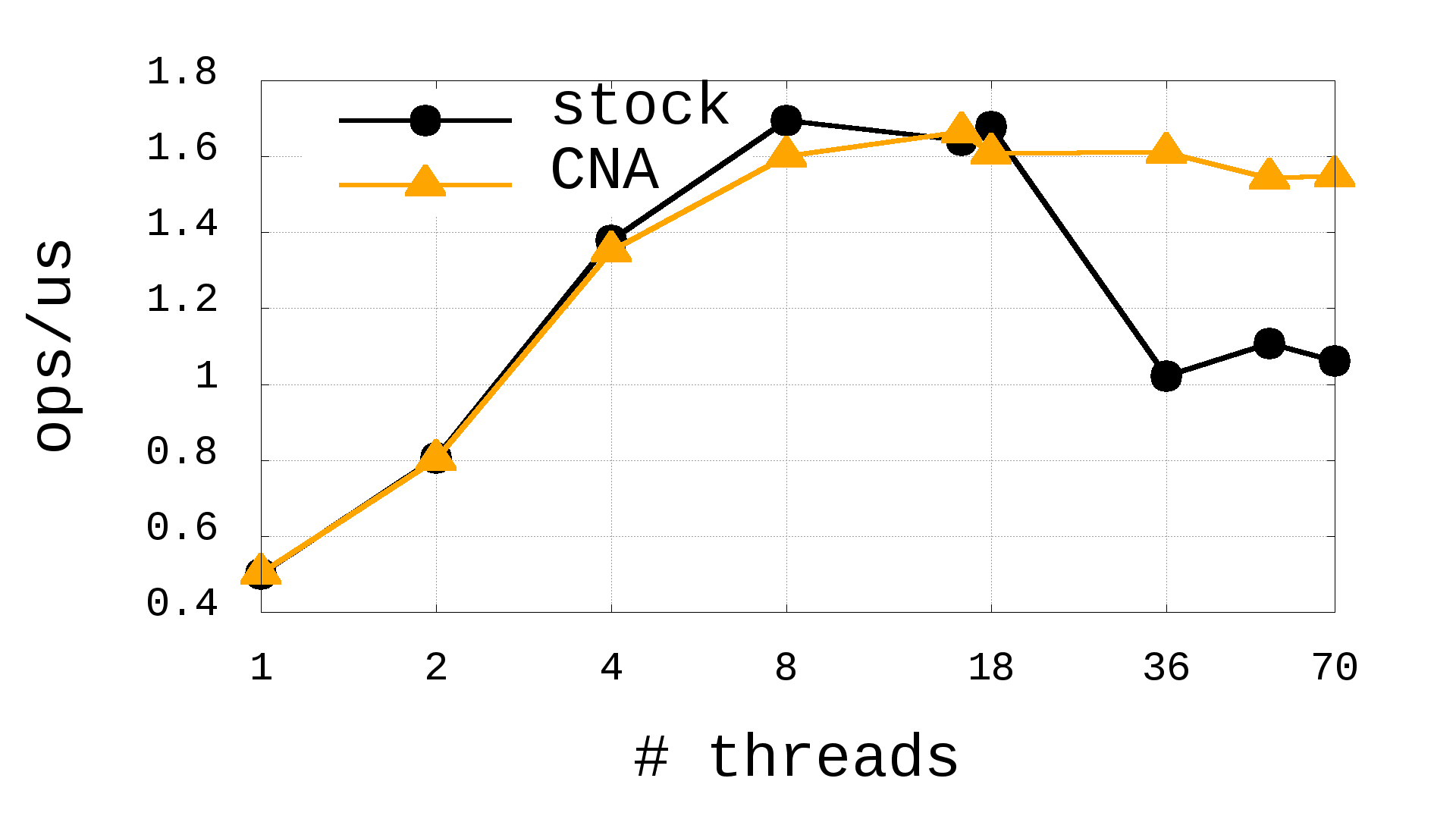}}
\subfloat[][open2\_threads]{\includegraphics[width=0.5\linewidth]{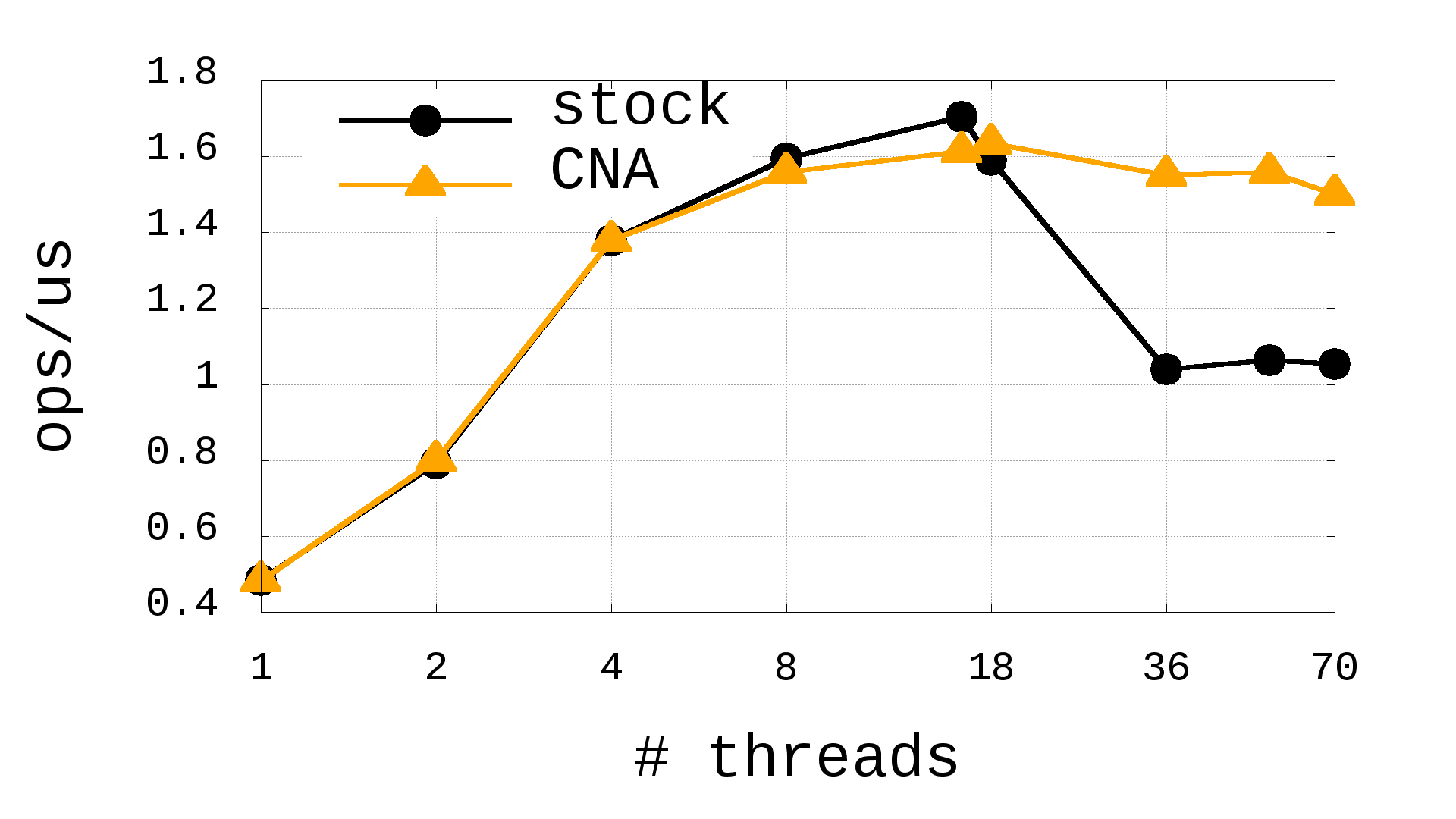}}
\caption{Performance results for the \code{will-it-scale} benchmarks.}
\figlabel{fig:will-it-scale-neelam}
\end{figure*}

\remove{
\begin{figure*}
\subfloat[][lock1\_threads]{\includegraphics[width=0.5\linewidth]{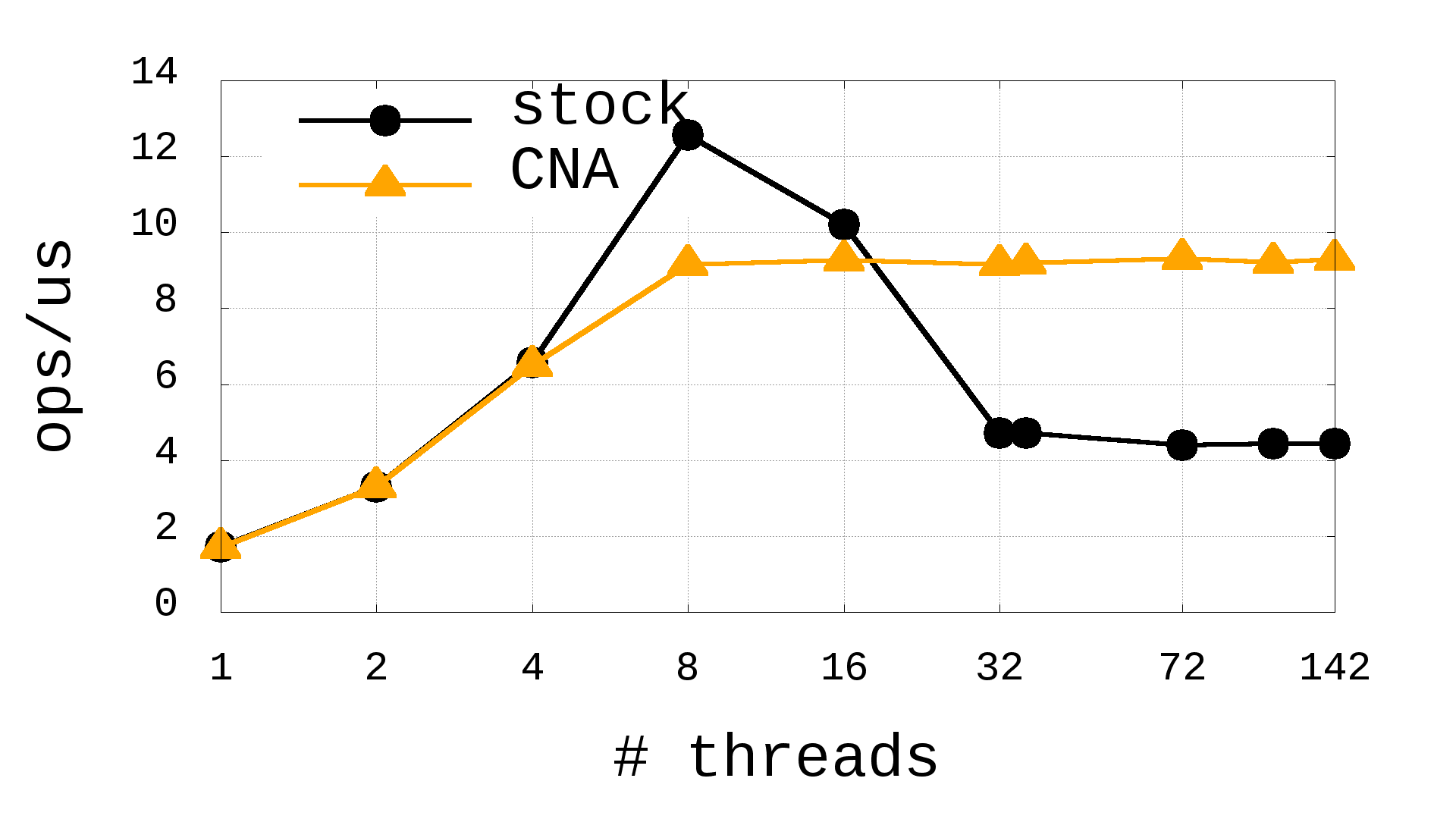}}
\subfloat[][lock2\_threads]{\includegraphics[width=0.5\linewidth]{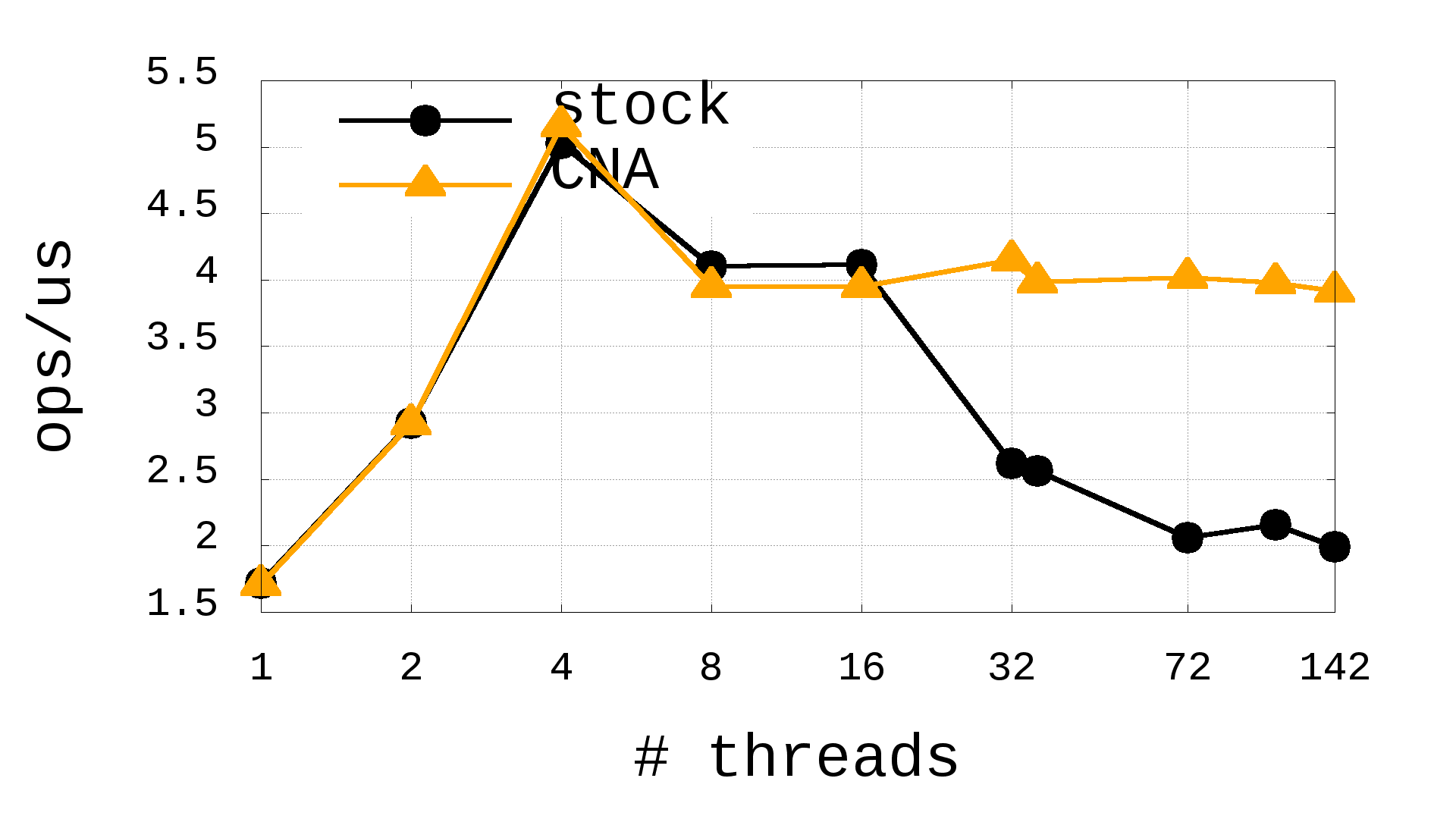}}\\
\subfloat[][open1\_threads]{\includegraphics[width=0.5\linewidth]{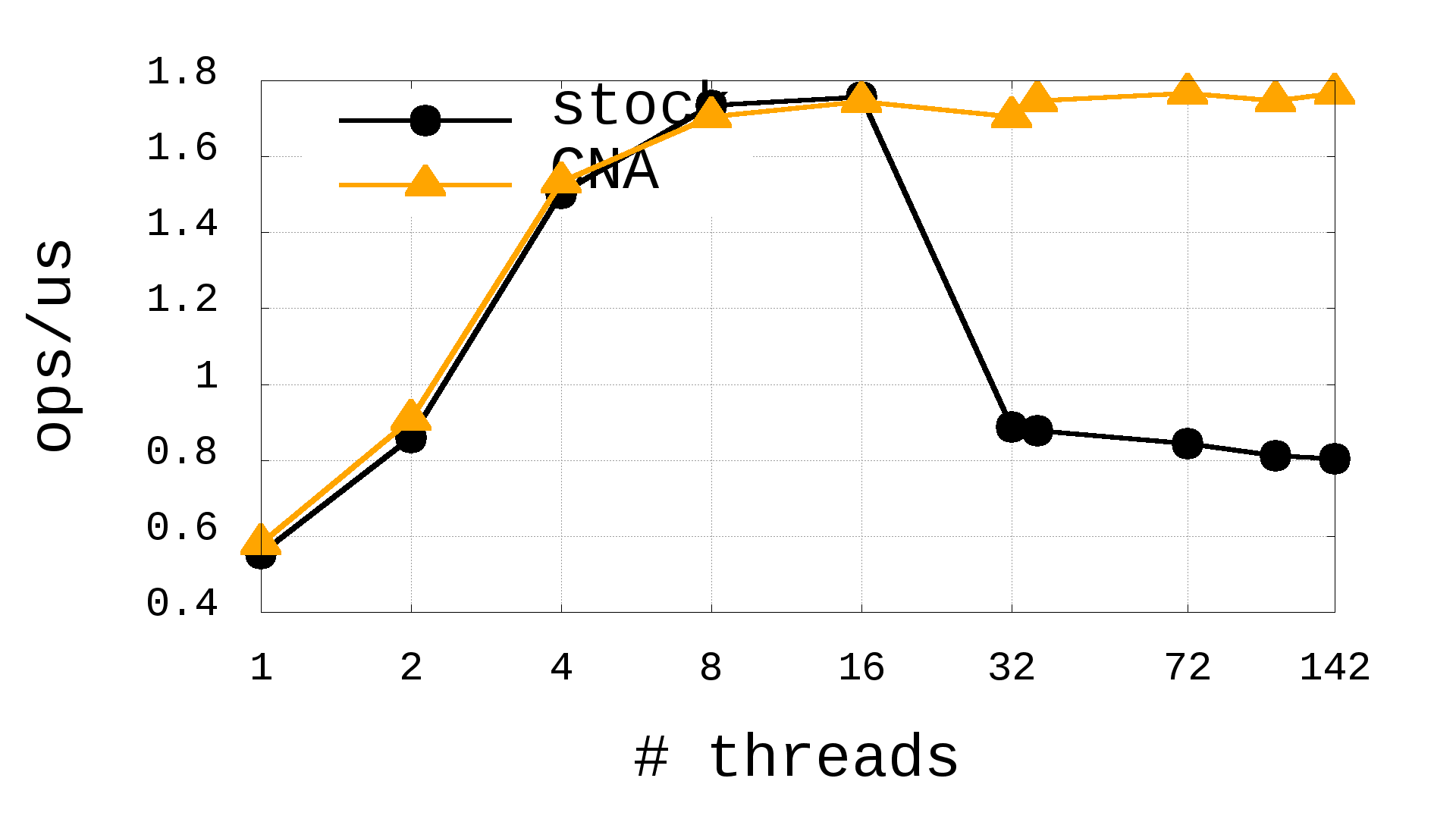}}
\subfloat[][open2\_threads]{\includegraphics[width=0.5\linewidth]{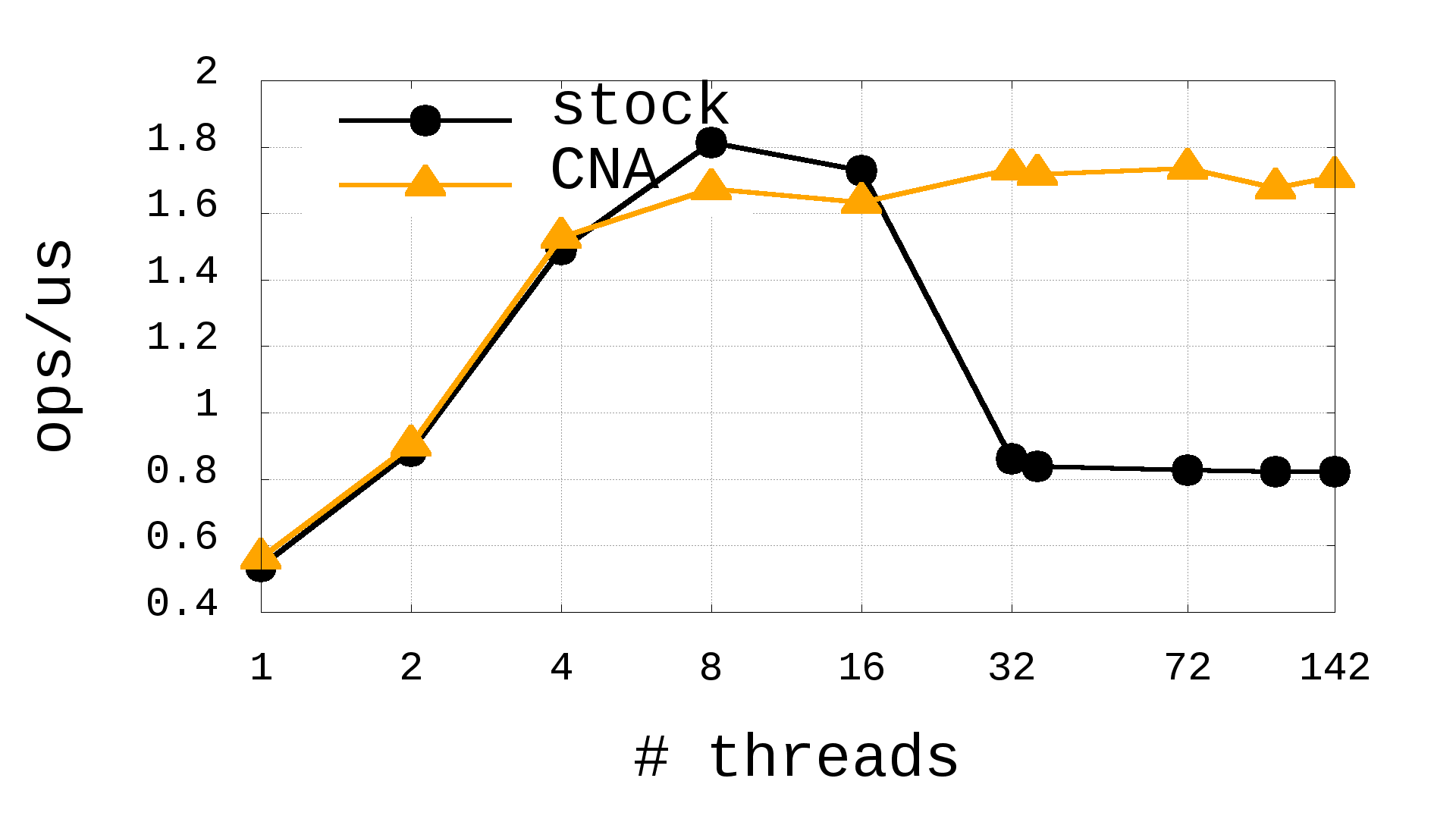}}
\caption{Performance results for the \code{will-it-scale} benchmarks.}
\figlabel{fig:will-it-scale-ol-bur-x5-4}
\end{figure*}
}

\subsubsection{will-it-scale}

Finally, we present the results of \code{will-it-scale}, a suite of user-space microbenchmarks designed 
to stress various kernel sub-systems.\footnote{https://github.com/antonblanchard/will-it-scale}
We experimented with several microbenchmarks in \code{will-it-scale}, and based on \code{lockstat} 
statistics identified a few that create contention on various spin locks in the kernel.
We note that for microbenchmarks that have not showed signs of spin lock contention in \code{lockstat},
the \code{CNA} version performed similarly to \code{stock}.
On the other hand, microbenchmarks with contention on one (or more) spin locks in the 
kernel exhibited a similar performance pattern, as demonstrated by a few examples included below.
We also note that we used \code{lockstat} only to identify benchmarks that create contention on a 
spin lock(s); performance numbers reported below were taken in the default configuration
with \code{lockstat} disabled, to avoid the probing effect the latter creates.

\figref{fig:will-it-scale-neelam} presents the results for four microbenchmarks from \code{will-it-scale}.
In the first pair, threads repeatedly lock and unlock a file lock through the \code{fcntl} command,
with the only difference that in \code{lock1\_threads} each thread works with a separate file, while
in \code{lock2\_threads} all threads operate on a lock associated with the same file.
In the second pair, threads repeatedly open and close a file (separate file for each thread)
in either the same directory (\code{open1\_threads}) or a different directory for each thread (\code{open2\_threads}).
The summary of points of contention in each of those benchmarks is given in Table~\ref{table:will-it-scale}.


Overall, the charts in \figref{fig:will-it-scale-neelam} show a behavior similar to the key-value 
map microbenchmark presented in \figref{fig:neelam-avl-tree-tput-with-extended-work}. 
Specifically, the \code{CNA} version matches the performance of \code{stock} as long as they both scale
(and where the contention on the spin locks detailed in Table~\ref{table:will-it-scale} does not exist yet).
At the performance peak, the \code{CNA} version slightly underperforms \code{stock} (by about 10\%)
as it pays for the overhead of restructuring the queue of waiting threads without any benefit.
Interestingly, the shuffle reduction optimization discussed above did not provide a relief in this case,
likely requiring more tuning, left for future work.
Along with that, as the contention on the respective spin lock(s) increases, the performance of \code{stock} degrades, while
the \code{CNA} version maintains a close-to-peak performance level.
This allows the \code{CNA} version to outperform \code{stock} by $42$--$57\%$ at $70$ threads. 

We note that like in all previous cases, the results on the four-socket machine were similar, 
with the dominance of \code{CNA} over \code{stock} even more pronounced.
For instance, \code{CNA} outperformed \code{stock} in \code{open1\_threads} by $120\%$ at $142$ threads.

\begin{table}
\resizebox{\linewidth}{!}{%
\begin{tabular}{|l|c|c|}
\hline
Benchmark & Contended spin locks & Call sites\\
\hline
\hline
\code{lock1\_threads} & \code{files\_struct.file\_lock} & \makecell{\code{\_\_alloc\_fd} \\ \code{fcntl\_setlk}}\\
\hline
\code{lock2\_threads} & \code{file\_lock\_context.flc\_lock} & \makecell{\code{posix\_lock\_inode}}\\
\hline
\multirow{2}{*}{\code{open1\_threads}} & \code{files\_struct.file\_lock} &  \makecell{\code{\_\_alloc\_fd} \\ \code{\_\_close\_fd}} \\\cline{2-3}
& \code{lockref.lock}  & \makecell{\code{dput} \\ \code{d\_alloc} \\ \code{lockref\_get\_not\_zero} \\ \code{lockref\_get\_not\_dead}}\\
\hline
\code{open2\_threads} & \code{files\_struct.file\_lock} & \makecell{\code{\_\_alloc\_fd} \\ \code{\_\_close\_fd}}\\
\hline
\end{tabular}
}
\caption{Contention in the \code{will-it-scale} benchmarks.}
\label{table:will-it-scale}
\end{table}

\section{Conclusion}
\seclabel{sec:conclusion}
The paper presents the construction of CNA, a compact NUMA-aware queue spin lock.
Unlike state-of-the-art NUMA-aware locks, which are hierarchical in their nature and thus have memory footprint size proportional to the number of sockets,
CNA's state requires only one word of memory.
This feature paired with the fact the CNA requires only one atomic instruction for acquisition (and at most one for release)
make CNA an attractive alternative for any (NUMA-oblivious or NUMA-aware) lock.

We implemented CNA as a stand-alone POSIX API-compliant library as well as integrated it 
into the Linux kernel, replacing the implementation of the spin lock slow path in the latter.
Our evaluation using both user-space and kernel benchmarks shows that CNA matches MCS in single-thread performance,
but outperforms it (typically, by about $40\%$ on a two-socket machine and about $100\%$ on a four-socket machine) under contention.
It is achieved by reducing the number of remote cache misses in lock handovers and in data accesses performed in 
critical sections protected by a lock.
At the same time, the admission policy of CNA achieves the long-term fairness comparable to that of MCS.
When compared to state-of-the-art NUMA-aware locks, CNA achieves a similar 
level of performance despite requiring substantially less space.

In the future work, we aim to explore further the effect of various optimizations described in this paper on the performance of CNA, and
also evaluate CNA in other user-space and kernel benchmarks.

\begin{acks}
We thank the anonymous reviewers as well as our shepherd Manuel Costa for valuable comments and suggestions to 
improve the presentation.
\end{acks}

\bibliography{refs}

\end{document}

%% file: defs.tex
\newcommand{\var}[1]{\lstinline+#1+}
\newcommand{\cSetP}[1]{\lstinline+Set<#1>+}

\newcommand{\figlabel}[1]{\label{figure:#1}}
\newcommand{\nakedfigref}[1]{\ref{figure:#1}}
\newcommand{\figref}[1]{Figure~\nakedfigref{#1}}

\newcommand{\linelabel}[1]{\label{line:#1}}
\newcommand{\nakedlineref}[1]{\ref{line:#1}}
\newcommand{\lineref}[1]{Line~\nakedlineref{#1}}
\newcommand{\linerangeref}[2]{Lines~\nakedlineref{#1}--\nakedlineref{#2}}

\newcommand{\seclabel}[1]{\label{sec:#1}}
\newcommand{\nakedsecref}[1]{\ref{sec:#1}}
\newcommand{\secref}[1]{Section~\nakedsecref{#1}}

\newcommand{\remove}[1] {}
\newcommand{\extabstract}[1] {}
\newcommand{\code}[1] {\texttt{#1}}